\documentclass[aps,prb,reprint,superscriptaddress]{revtex4-2}
\usepackage[bookmarks=true,colorlinks=true,urlcolor=blue,linkcolor=blue,citecolor=blue,breaklinks]{hyperref}
\usepackage{amsmath, amstext, amssymb, amsfonts, amsxtra}
\usepackage[table]{xcolor}
\usepackage{multirow}
\usepackage{grffile}
\usepackage{bm}
\usepackage{makecell}
\newcommand{\xiamen}{Department of Physics and Fujian Provincial Key Laboratory of Low Dimensional Condensed Matter Physics, Xiamen University, Xiamen 361005, Fujian, China}
\newcommand{\lanzhou}{Lanzhou Center for Theoretical Physics, Lanzhou University, Lanzhou 730000, Gansu, China}

\begin{document}

\title{Dynamics of wavepackets and entanglement in many-body kicked rotors under quantum resonance}

\author{Yangshuo Zhou}
\affiliation{\xiamen}
\author{Jiao Wang}
\email{phywangj@xmu.edu.cn}
\affiliation{\xiamen}
\affiliation{\lanzhou}


\begin{abstract}
We investigate a many-body interacting system of quantum kicked rotors, where each rotor resides in its respective quantum resonance. Rich many-body dynamics are found to emerge from the interplay between the principal and secondary resonances. In particular, for both the wavepacket and bipartite entanglement entropy, we analytically demonstrate three distinct dynamical regimes—quadratic spreading (growth), period-2 oscillation, and their hybrid—governed by the respective symmetries of the relevant potentials. Based on these symmetries, the connection between the wavepacket and the entanglement dynamics is illustrated. Other related issues are also discussed, including higher-order resonance effects, the robustness of the predicted dynamical behaviors, extension to many-body kicked tops, and relevance to experimental studies.
\end{abstract}
\maketitle

\section{Introduction}

Entanglement, as one of the most intriguing quantum effects, has stimulated enormous research interest, driving advances in quantum information science and cutting-edge technologies such as quantum computing. Entanglement-related studies have also provided new insights into other fundamental issues including thermalization, many-body localization, and the quantum–classical correspondence \cite{Abanin2019, Kaufman2016, Jacquod2009, Calabrese2005, Lerose2020, Laksh2016}.

Entanglement dynamics is a vibrant subfield that explores how entanglement is generated, evolves, and spreads during the unitary evolution of a composite system. A common setup considers an initially disentangled localized wavepacket, motivated partly by theoretical tractability and experimental feasibility. A key line of research aims to connect entanglement dynamics with wavepacket dynamics, to examine whether hierarchical structures established for the latter can be translated to the former. Important progress has been made in this regard. For example, linear and logarithmic entanglement entropy growth on the Ehrenfest time scale has been linked to underlying chaotic and integrable dynamics, respectively \cite{Lerose2020}.

The two-body quantum kicked rotors (QKRs) have served as a central model in this direction \cite{Adachi1988, Doron1988, Gadway2013, Matsui2016, Notar2018, Rozenbaum2017, Laksh2016, SP2020, JJP2020, AN2024, Fujisaki2003, SP2024}. As the simplest extension of the single-body QKR---a paradigmatic model for quantum chaos \cite{Casati1979, Fishman1982, Izrailev1990, Casati1995} relevant to diverse fundamental problems \cite{Garreau2017}, it inherits the latter's simplicity and richness. Studies using the two-body QKRs have clarified several key issues, including the finding that stronger classical chaos does not necessarily enhance entanglement growth \cite{Fujisaki2003}.

Despite these advances, a general understanding of why wavepacket and entanglement dynamics can be connected remains lacking. For a bipartite many-body system, entanglement dynamics is governed solely by the inter-subsystem interaction. By contrast, wavepacket dynamics depends on both inter- and intra-subsystem interactions and thus cannot be fully reconstructed from entanglement alone. Conversely, bipartition is an external, artificial choice whose information is not encoded in the system Hamiltonian. Neither inter-subsystem interaction nor entanglement dynamics can therefore be determined generically from wavepacket dynamics.  Then, to what extent and through what mechanisms can wavepacket and entanglement dynamics be connected?

In this work, our aim is to revisit this issue under an unexplored model---a general in-resonance many-body QKR model where each rotor is tuned to a quantum resonance.  Quantum resonance in the single-body QKR occurs when a control paramter becomes a rational multiple of $4\pi$, with the two lowest-order resonances at $4\pi$ and $2\pi$ referred to as principal and secondary resonances, respectively. In contrast to the dynamical localization of off-resonance rotors \cite{Casati1979, Fishman1982, Izrailev1990, Casati1995}, in-resonance wavepackets typically exhibit unbounded spreading. Both regimes carry fundamental significance \cite{Garreau2017}.  Yet the influence of quantum resonance on wavepacket dynamics in many-body QKRs remains largely unexplorted. Similarly,  its effect on entanglement dynamics has been little studied, with only one work that considers the special case in which two or three rotors are all set at the principal resonance \cite{SP2024}. Most importantly, whether and how wavepacket and entanglement dynamics are linked under quantum resonance has never been examined. Our work fills these gaps.

We find that indeed, quantum resonance induces rich and intricate wavepacket and entanglement dynamical behaviors, governed by the symmetry nature of each rotor's effective potential and that of the inter-subsystem interacion.  Wavepacket and entanglement dynamics are not linked directly, but are subject to the ``selection rule'' imposed by the underlying symmetries, allowing one to be inferred from the other in some cases.

This paper is organized as follows. In Sec. II, we review quantum resonance in the single-body QKR. In Sec. III, we derive analytical results for wavepacket and entanglement dynamics in many-body QKRs at the two lowest-order resonances and discuss their connection, supported by numerical evidence in Sec. IV. Higher-order resonances are studied in Sec. V, and the robustness of our findings is analyzed in Sec. VI. In Sec. VII, we generalize our results to many-body quantum kicked tops. Finally, Sec. VIII provides a summary and outlook.

\section{Single-body kicked rotor}
The conventional single-body kicked rotor serves as the building block of our many-body interacting system. In dimensionless units, its Hamiltonian is given by
\begin{equation}
H = \frac{\tau {p}^2}{2} + V(\theta) \sum_n \delta(t-n T).
\end{equation}
Here, $1/\tau$ represents the rotational inertia of the rotor, and $\theta$ and $p$ are, respectively, its angular coordinate and the corresponding conjugate angular momentum. The rotor is kicked instantaneously and periodically with period $T$ and kicking potential $V(\theta)$, and in between two consecutive kicks, the rotor rotates freely. For the standard QKR, $V(\theta)=k\cos(\theta)$, where $k$ is an overall kick strength. But note that in what follows, we assume a general  $V(\theta)$ that is not restricted to this specific form.

The wavepacket dynamics of the QKR is qualitatively governed by the parameter $\alpha=\tau T \hbar$ \cite{Casati1979, Fishman1982, Izrailev1990, Casati1995}. Hence, for convenience and without loss of generality, we take the units in which $T=\hbar=1$ and thus omit them from the subsequent expressions, leaving $\tau$ the only relevant parameter. Consequently, the eigenvalues of the angular momentum $p$ are integers and we denote the eigenstate of eigenvalue $l$ by $|l\rangle$; i.e., $p|l\rangle=l|l\rangle$, for $l\in \mathbb Z$. The evolution operator $U$ over a period thus reads
\begin{equation}
U =  \exp\left(-i\frac{\tau p^2}{2}\right) \exp\left(-iV(\theta)\right).
\end{equation}
The former operator $U_f = \exp(-i\tau p^2/2)$ accounts for the free rotation, while the latter $U_k=\exp (-iV(\theta))$ describes the kicking process. 

For a general value of $\tau$ (i.e., $\alpha$), because of destructive interference, a wavepacket typically ceases spreading in the angular momentum space at a crossover time, which is known as ``the dynamical localization'' \cite{Casati1979, Fishman1982, Izrailev1990, Casati1995}. However, when $\tau$ is a rational multiple of $\pi$, the operator $U$ possesses a discrete translational symmetry. Hence, in general under this condition, by Bloch's theorem, a wavepacket keeps spreading such that the variance of the angular momentum increases quadratically in time. This is known as ``the quantum resonance" \cite{Izrailev1980}. Both regimes are remarkable fingerprints of the QKR.

For $\tau=4\pi r/s$ with coprime integers $r$ and $s$, the system is in an $s$-order quantum resonance. The lowest order resonance ($s=1$) is referred to as the principal resonance, where the quadratic wavepacket spreading can be conveniently appreciated. In this case, the operator for free rotation reduces to the identity, $U_f= \mathbb{I}$, hence the evolution operator for $t$ periods, or $t$ steps, becomes $U^t=U_k^t=\exp (-itV(\theta))$. Taking for example the eigenstate $|0\rangle$ of $p$ with 0 eigenvalue as the initial state, it follows immediately that the variance in the angular momentum (or, equivalently, the kinetic energy) grows quadratically as $\langle p^2(t) \rangle = \langle 0|{U^t}^{\dagger}  p^2 U^t |0\rangle=\lambda t^2$ with
\begin{equation}
\lambda=\frac1{2\pi}\int d\theta \left|\frac{\partial  V}{\partial\theta}\right|^2.
\label{psq}
\end{equation}
Here, $\sqrt{\lambda}$ represents the mean free path of the kick operator~\cite{Doron1988}, quantifying the momentum imparted by a kick. Such a ballistic spreading is a generic feature for almost all quantum resonances but one.

The only exception occurs for the secondary resonance with $s=2$ such that $\tau=2\pi$. In this case, 
the free rotation operator reduces to $U_f = \exp({-i{\tau p^2}/{2}})=\exp({-i{\pi p^2}})=\exp({-i{\pi p}})$ 
\footnote{Applying the two operators $\exp({-i{\pi p^2}})$ and $\exp({-i{\pi p}})$ to an eigenstate $|l\rangle$ of $p$ leads to the same result, i.e., $\exp(-il\pi)|l\rangle=(-1)^l |l\rangle$, because of the fact that an integer and its square share the same parity. It follows straightforwardly that the two operators are equivalent.}, 
which is a translation operator in $\theta$ that translates the coordinate $\theta$ by $\pi$. As a consequence,
\begin{equation}
 U_{f}\exp \left(-iV(\theta)\right)=\exp\left(-iV(\theta +\pi)\right)U_{f}.
\end{equation}
The evolution operator over two periods then reads $U^2=U_fU_kU_fU_k = \exp(-i[V(\theta)+V(\theta+\pi)])$, considering that $U_f^2 = \mathbb I$.

Obviously, the dynamics of the QKR depends critically on the symmetry of the potential $V(\theta)$. If the potential is antisymmetric under a $\pi$-shift, i.e., $V(\theta +\pi)=-V(\theta)$---as is the standard QKR
with $V(\theta) = k\cos(\theta)$---the two-step evolution operator reduces to the identity, \(U^2 = \mathbb I\). The wavepacket spreading is thus suppressed and its dynamics displays a period-2 oscillation, which is termed ``quantum anti-resonance'' to highlight its distinction from the general quantum resonance \cite{Izrailev1980, Dana1996}.

Conversely, if the potential is symmetric with respect to a $\pi$-shift, $V(\theta +\pi)=V(\theta)$, the two-step evolution operator becomes $U^2 = \exp\left(-i2V(\theta)\right)$ instead, which collapses to that in the principal resonance. In this case, the wavepacket spreads ballistically, $\langle p^2(t) \rangle =\lambda t^2$, with $\lambda$ being given by Eq.~\eqref{psq} as well.

For an asymmetric potential, it can be decompose into a superposition of antisymmetric and symmetric components. Consequently, the wavepacket dynamics is dominated by ballistic spreading from the symmetric component, with a period-2 oscillatory modulation arising from the antisymmetric component.

The knowledge of quantum resonance in the single-body QKR lays the groundwork for our analysis of the many-body QKRs in the following sections.

\section{Many-body kicked rotors}
In this section, we discuss the wavepacket and entanglement dynamics in many-body interacting kicked rotors. The Hamiltonian of a general $N$-body kicked rotor system can be written as
\begin{equation}
	H=\sum_{j = 1}^N \frac{\tau_j p_j^2}{2}+ V(\theta_1,...,\theta_N)\sum_n\delta(t-n),
	\label{Hamiltonian}
\end{equation}
where $\theta_j$ and $p_j$ are the conjugate angular coordinate and momentum of the $j$-th rotor, respectively, with rotational inertia $1/\tau_j$. For brevity, we will hereafter use the vector notation $\boldsymbol{\theta}\equiv(\theta_1,..., \theta_N)$ and ${\mathbf{p}}\equiv(p_1,..., p_N)$  to denote the sets of all rotor coordinates and momenta, respectively.

The evolution of the system is governed by the evolution operator
\begin{equation}
	U = \exp\left(-i\sum_{j = 1}^N \frac{\tau_j p_j^2}{2}\right)\exp\left(-iV(\boldsymbol{\theta})\right).
	\label{U1}
\end{equation}
To quantify the bipartite entanglement between a subsystem $A$ and its complement $B$, we adopt the linear entropy, $S_{\text{lin}}$, in view of its theoretical simplicity and experimental accessibility \cite{Islam2015, Ho2017}. For a given pure state of the entire composite system,  \(|\psi\rangle\), it is defined as
\begin{equation}
S_\text{lin} = 1-\mathrm{Tr} \left( {\rho}^2_A \right),
\end{equation}
where $ \rho_A\equiv\mathrm{Tr}_B(|\psi\rangle\langle\psi|)$ is the reduced density matrix of subsystem $A$.

As our motivation is to investigate the effects of quantum resonance, we place each rotor at a quantum resonance by setting $\tau_j = 4\pi r_j/s_j$, where $r_j$ and $s_j$ are two coprime integers corresponding to the $s_j$-order quantum resonance. For the sake of convenience, we postpone the discussions of higher order resonances to Sec. V, but restrict ourselves to the principal and secondary resonances in the rest of this section.

We now construct our analytical framework. Suppose that in the system, \(q\) out of the \(N\) rotors are at the secondary resonance, while the remaining \(N-q\) are at the principal resonance, such that
\begin{equation}
\tau_j=
    \begin{cases}
        2\pi, &j \in \mathcal S ,\\
        4\pi, &\text{otherwise},
    \end{cases}
\end{equation}
where \(\mathcal{S}\equiv \{j_1, \dots, j_q\}\) is the set of indices for the rotors at the secondary resonance.
This specific choice of resonance condition simplifies the analysis significantly: as shown in Sec. II, the free rotation operator for any principal-resonance rotor reduces to \(\exp(-i4\pi p_j^2/2) = \ \mathbb{I}\), having no net effect on the rotor's state, while that for secondary-resonance rotors becomes the translational operator \(\exp(-i2\pi  p_j^2/2) = \exp(-i\pi p_j)\). The evolution operator thus reads
\begin{equation}
\begin{aligned}
    U &= U_{\pi}\exp\left(-iV(\boldsymbol{\theta})\right),
\end{aligned}
\label{U12}
\end{equation}
where the free rotation operator
\begin{equation}
     U_{\pi}\equiv\exp\left[{-i\pi\sum_{j\in \mathcal{S}} p_{j}}\right]
     \label{Upi}
\end{equation}
generates a translation that shifts the angular coordinate by \(\pi\) for each rotor in \(\mathcal S\) but keeps the angular coordinate unchanged for others.  We have therefore the commutation relation
\begin{equation}
    U_{\pi}\exp(-iV(\boldsymbol{\theta}))=\exp(-iV(\boldsymbol{\theta'}))U_{\pi}
\end{equation}
with \(\boldsymbol{\theta'}\equiv (\theta_1',\dots,\theta_N')\) being the shifted coordinates that
\begin{equation}
\theta_j'=
    \begin{cases}
        \theta_j+\pi,&j\in \mathcal{S} ,\\
        \theta_j,&\text{otherwise}.
    \end{cases}
    \label{thetaprime}
\end{equation}

The algebraic structure of this commutation relation allows for the factorization of the \( t \)-step evolution operator into a purely momentum-dependent term and a purely coordinate-dependent term,
\begin{equation}
U^t = \exp(-i\mathcal{K}_t({\mathbf{p}}))\exp(-i\mathcal{V}_t(\boldsymbol{\theta})),
\label{Ut}
\end{equation}
a key simplification that facilitates the subsequent analysis. Here, \(\mathcal{K}_t ({\mathbf{p}})\) and \(\mathcal{V}_t (\boldsymbol{\theta})\) are, respectively, the effective accumulated kinetic and potential energy. To demonstrate this, note that the two-step evolution operator reduces
\begin{equation}
    U^2 = \exp\left(-i[V(\boldsymbol{\theta})+V(\boldsymbol{\theta'})]\right),\\
\label{u2rotor}
\end{equation}
since the partial \(\pi\)-translation operator $U_{\pi}$ satisfies \( U_{\pi}^2=\mathbb{I}\).
The evolution operator for all even-time steps (\(t =2m\)) follows immediately,
\begin{equation}
U^{2m} = \exp\left(-im[V(\boldsymbol{\theta})+V(\boldsymbol{\theta'})]\right),
\label{U2m}
\end{equation}
and in turn, that for odd-time steps (\(t =2m+1\)) reads
\begin{equation}
U^{2m+1} = U_{\pi}\exp\left(-i[(m+1)V(\boldsymbol{\theta})+mV(\boldsymbol{\theta'})]\right).
\label{U2m1}
\end{equation}
Then, by collecting the coordinate- and momentum-dependent terms from Eqs. \eqref{U2m} and \eqref{U2m1}, we can read off the explicit forms of the effective accumulated kinetic energy,
\begin{equation}
\begin{aligned}
        \mathcal{K}_t ({\mathbf{p}}) &=\begin{cases}
        0, &t = 2m,\\
        \pi\sum_{j\in \mathcal{S}} p_{j}, &t = 2m+1,
    \end{cases}\\
    \end{aligned}
\end{equation}
and the effective accumulated potential energy,
\begin{equation}
\begin{aligned}
    \mathcal{V}_{t} (\boldsymbol{\theta}) & = \begin{cases}
        m(V(\boldsymbol{\theta})+V(\boldsymbol{\theta'})), &t = 2m,\\
        (m+1)V(\boldsymbol{\theta})+mV(\boldsymbol{\theta'}), &t = 2m+1.
         \end{cases}
         \label{Vt}
\end{aligned}
\end{equation}

Now, we are ready to investigate the wavepacket and the entanglement dynamics. 
For the wavepacket, we use its mean and mean-squared displacement in each rotor’s momentum space to characterize its dynamics. For any given initial state \(|\psi(0)\rangle\), after $t$ evolution steps, these quantities are given by
\begin{equation}
\begin{aligned}
      D_j(t)&\equiv \langle \psi(0)|p_j (t)-p_j(0) |\psi(0)\rangle\\
      &=\langle \psi(0) |{U^t}^{\dagger} p_j (0) U^t -p_j(0)|\psi(0)\rangle\\
      &=\int  d\theta_1 \dots d\theta_N |\langle \boldsymbol{\theta}|\psi(0)\rangle|^2
      \left(-\frac{\partial \mathcal V_t}{\partial\theta_j}\right)\\
\end{aligned}
\end{equation}
and
\begin{equation}
\begin{aligned}
      \sigma_j^2(t)&\equiv \langle \psi(0)| [p_j (t)-p_j(0)]^2 |\psi(0)\rangle\\
      &=\langle \psi(0) |[{U^t}^{\dagger} p_j (0) U^t -p_j(0)]^2|\psi(0)\rangle\\
      &=\int  d\theta_1 \dots d\theta_N |\langle \boldsymbol{\theta}|\psi(0)\rangle|^2
     \left[\frac{\partial \mathcal V_t}{\partial\theta_j}\right]^2.\\
\end{aligned}
\end{equation}
While \(D_j(t)\) captures the overall drift of the wavepacket in momentum space, \(\sigma_j^2(t)\) (togather with \(D_j(t)\)) captures its spreading: for a wavepacket initially localized in momentum spaces, \(\text{Var}(p_j(t)-p_j(0)) = \sigma_j^2(t) - D_j(t)^2\) quantifies the broadening of the wavepacket during evolution.

Obviously, both $D_j(t)$ and $\sigma_j^2(t)$ are governed by the properties of $\partial \mathcal{V}_t/\partial \theta_j$ and in turn, by the properties of $\partial V/\partial \theta_j$ owing to Eq. \eqref{Vt}. As only the terms in  $V(\boldsymbol{\theta})$ that involve $\theta_j$ are relevant to $\partial V/\partial \theta_j$, for convenience in the following discussion, we define an effective potential for the $j$-th rotor, denoted as $V_j(\boldsymbol{\theta})$, as the one that retains only these terms, excluding all others independent of $\theta_j$.

To appreciate the role played by $V_j(\boldsymbol{\theta})$, we note that the \(\pi\)-translation operator is an involution (\( U_{\pi}^2=\mathbb{I}\)), whose eigenvalues are restricted to \(\pm 1\). This allows $V_j(\boldsymbol{\theta})$ to be uniquely decomposed into two components that correspond to these two eigenspaces, respectively. Namely, a symmetric component \(V_{j+}(\boldsymbol{\theta})\equiv [V_{j}(\boldsymbol{\theta})+V_{j}(\boldsymbol{\theta'})]/2\) with \(V_{j+}(\boldsymbol{\theta'})=V_{j+}(\boldsymbol{\theta})\) and an antisymmetric component \(V_{j-}(\boldsymbol{\theta})\equiv [V_{j}(\boldsymbol{\theta})-V_{j}(\boldsymbol{\theta'})]/2\) with \(V_{j-}(\boldsymbol{\theta'})=-V_{j-}(\boldsymbol{\theta})\). The effect of each component on the wavepacket dynamics can therefore be quantified with the following parameters:
\begin{equation}
\alpha_{j\pm} \equiv \left\langle  -\frac{\partial V_{j\pm}(\boldsymbol{\theta})}{\partial\theta_j}\right\rangle,
\label{eq:l_alpha}
\end{equation}
\begin{equation}
\lambda_{j\pm}\equiv \left\langle \left[\frac{\partial V_{j\pm}(\boldsymbol{\theta})}{\partial\theta_j}\right]^2\right\rangle,
\label{eq:l_lambda}
\end{equation}
and
\begin{equation}
\kappa_j\equiv \left\langle  
\frac{\partial V_{j+}(\boldsymbol{\theta})}{\partial\theta_j}\frac{\partial V_{j-}(\boldsymbol{\theta})}{\partial\theta_j}\right\rangle,
\label{eq:l_kappa}
\end{equation}
where the angular bracket \(\langle\cdot\rangle\) denotes the average over the initial state, i.e., for a given function $f(\boldsymbol{\theta})$,
\begin{equation}
\langle f(\boldsymbol{\theta})\rangle \equiv \int d\theta_1 \dots d\theta_N |\langle \boldsymbol{\theta} |\psi(0)\rangle|^2 
f(\boldsymbol{\theta}).
\label{eq:l_kappa}
\end{equation}

First of all, if the potential is purely symmetric, i.e., \(V_{j-}(\boldsymbol{\theta}) = 0\) and $V_j(\boldsymbol{\theta}) = V_{j+}(\boldsymbol{\theta})$, then $\partial \mathcal{V}_t/\partial \theta_j = t \partial V_{j+}/\partial\theta_j$, and the parameters for the antisymmetric and cross terms vanish identically, \(\alpha_{j-} =\lambda_{j-} =\kappa_j = 0\), leading to \(D_j(t) = t \alpha_{j+}\) and \(\sigma_j^2(t) = t^2 \lambda_{j+}\). It suggests that, in general, the wavepacket may drift  (provided $\alpha_{j+}\ne 0$)  and spread quadratically if $\lambda_{j+}> \alpha_{j+}^2$, a condition easily satisfied as $\lambda_{j+}\ge \alpha_{j+}^2$ by definition. 

In contrast, if the potential is antisymmetric, i.e.,  $V_{j+}(\boldsymbol{\theta}) = 0$ and $V_j(\boldsymbol{\theta}) = V_{j-}(\boldsymbol{\theta})$, then we have $\alpha_{j+}  =\lambda_{j+} =\kappa_j = 0$, suggesting that both 
$D_j$ and $\sigma_j^2$ oscillate between zero and a constant: \(D_j=0\) at even-time steps and \(D_j=\alpha_{j-}\) at odd-time steps; \(\sigma_j^2=0\) at even-time steps and \(\sigma_j^2=\lambda_{j-}\) at odd-time steps.

In the general case in which $V_j(\boldsymbol{\theta})$ is asymmetric, the wavepacket dynamics exhibits a hybrid behavior of that in the symmetric and antisymmetric cases. The mean displacement $D_j$ is dominated by the linear growth originating from $V_{j+}(\boldsymbol{\theta})$, which is modulated by the constant offset from  $V_{j-}(\boldsymbol{\theta})$ at odd-time steps:
\begin{equation}
 D_j(t) = \begin{cases}t \alpha_{j+}, & t = 2m, \\ t\alpha_{j+} + \alpha_{j-}, & t = 2m+1.\end{cases}
\end{equation}
The mean squared displacement \(\sigma_j^2\) behaves similarly: it is primarily quadratic, but with linear and constant offsets every two steps:
\begin{equation}\sigma_j^2(t) = 
\begin{cases}t^2 \lambda_{j+}, & t = 2m, \\
t^2 \lambda_{j+} + 2t \kappa_j + \lambda_{j-}, & t = 2m+1.
\end{cases}
\label{msd}
\end{equation}

Next, we turn to analyze the entanglement generation. We partition the whole system into two subsystems, \(A\) and \(B\), with the first \(N_A\) rotors assigned to \(A\) and the remaining \(N_B=N-N_A\) rotors to \(B\). The corresponding coordinate vectors are \(\boldsymbol{\theta}_A\) and \(\boldsymbol{\theta}_B\), respectively. Accordingly, the potential of the system can be partitioned into 
\(V(\boldsymbol{\theta}) = V_A(\boldsymbol{\theta}_A)+V_B(\boldsymbol{\theta}_B)+V_I(\boldsymbol{\theta}_A,\boldsymbol{\theta}_B)\), where \(V_A\) and \(V_B\) are ``local'' potentials only acting on \(A\) and \(B\), respectively, and \(V_I\) is the interaction potential that couples the two subsystems. For brevity, we denote \(V_I(\boldsymbol{\theta}_A, \boldsymbol{\theta}_B)\) by \(V_I(\boldsymbol{\theta})\) when no confusion arises, with \(\boldsymbol{\theta} \equiv (\boldsymbol{\theta}_A, \boldsymbol{\theta}_B)\).
Note that the indices of rotors at the secondary resonance, $j\in \mathcal{S}$, are arbitrarily distributed from 1 to $N$, and thus we can assign these secondary-resonance rotors (and hence those principal-resonance rotors complementarily) to $A$ and $B$ arbitrarily by setting the values of indices in $\mathcal{S}$.

The interaction \(V_I\) is necessary to entanglement generation, as local dynamics cannot establish correlations between subsystems~\cite{Wang2002}. For the case we are analyzing here, the \(t\)-step evolution operator (Eq.~\eqref{U2m} and ~\eqref{U2m1}) naturally factorizes as \(U^t=U_\text{local}^tU_I^t\), where the former local part \(U_\text{local}^t\) contains only the kinetic term and the local potentials \(V_A\) and \(V_B\), hence making no contribution to entanglement generation. Entanglement growth is therefore encoded in \(U_I^t\equiv \exp(-i\mathcal{V}_{t,I}(\boldsymbol\theta))\) exclusively with the corresponding effective accumulated interaction
\begin{equation}
    \mathcal{V}_{t,I}(\boldsymbol\theta) = \begin{cases}
        tV_{I+}(\boldsymbol{\theta}), &t = 2m,\\
         tV_{I+}(\boldsymbol{\theta})+V_{I-}(\boldsymbol{\theta}), &t = 2m+1,
    \end{cases}
    \label{VST1}
\end{equation}
where \(V_{I\pm}(\boldsymbol{\theta})\equiv [V_{I}(\boldsymbol{\theta})\pm V_{I}(\boldsymbol{\theta'})]/2\) represent the symmetric $(+)$ and antisymmetric $(-)$ parts of the interaction potential with respect to the partial \(\pi\)-translation.

We consider a generic pure disentangled initial state, i.e., a bipartite product state.  As the system evolves, however, the two subsystems may become entangled owing to interaction. Crucially, the bipartite linear entropy at any given time step $t$ can be derived explicitly (see Appendix A), a core result that enables us to uncover the underlying mechanism of entanglement generation. It reads
\begin{equation}
    S_{\text{lin}}(t)=1-\left\langle \cos(\Delta_t)\right\rangle.
    \label{SlintDelta}
\end{equation}
Here $\left\langle \cos(\Delta_t)\right\rangle$ represents the average of $\cos(\Delta_t)$ over the initial state (see Eq. \eqref{average} in Appendix A for its expression), where the phase \(\Delta_t(\boldsymbol{\theta}_A, \boldsymbol{\theta}_B, \boldsymbol{\theta}_{A'}, \boldsymbol{\theta}_{B'})\equiv \mathcal{V}_{t,I}(\boldsymbol{\theta}_A, \boldsymbol{\theta}_B) + \mathcal{V}_{t,I}(\boldsymbol{\theta}_{A'}, \boldsymbol{\theta}_{B'})- \mathcal{V}_{t,I}(\boldsymbol{\theta}_{A'}, \boldsymbol{\theta}_B) - \mathcal{V}_{t,I}(\boldsymbol{\theta}_A, \boldsymbol{\theta}_{B'})\)
records the energy differences between the two sets of effective accumulated potentials, where
$\boldsymbol{\theta}_{A'}$ and $\boldsymbol{\theta}_{B'}$ are two coordinate vectors for subsystem $A$ and $B$, respectively, independent of $\boldsymbol{\theta}_{A}$ and $\boldsymbol{\theta}_{B}$.
By substituting Eq.~\eqref{VST1} into the definition of \(\Delta_t\) and taking advantage of its linear structure, we can decompose the phase into symmetric and anti-symmetric contributions as well. This yields the desired result
\begin{equation}
    S_{\text{lin}}(t) = 
    \begin{cases}
        1 - \left\langle \cos(t\epsilon_+) \right\rangle, & t = 2m, \\[6pt]
        \begin{aligned}
            1 &- \left\langle \cos(t\epsilon_+) \cos(\epsilon_-) \right\rangle \\
              &+ \left\langle \sin(t\epsilon_+) \sin(\epsilon_-) \right\rangle,
        \end{aligned} & t = 2m+1.
    \end{cases}
    \label{ST}
\end{equation}
Here, \(\epsilon_{\pm}\), defined similarly as \(\Delta_t\) but for the energy differences of ${V}_{I\pm}$ instead, denotes \(\epsilon_{\pm}(\boldsymbol{\theta}_A, \boldsymbol{\theta}_B, \boldsymbol{\theta}_{A'}, \boldsymbol{\theta}_{B'})\equiv {V}_{I\pm}(\boldsymbol{\theta}_A, \boldsymbol{\theta}_B)+{V}_{I\pm}(\boldsymbol{\theta}_{A'}, \boldsymbol{\theta}_{B'})-{V}_{I\pm}(\boldsymbol{\theta}_{A'}, \boldsymbol{\theta}_B)-{V}_{I\pm}(\boldsymbol{\theta}_A, \boldsymbol{\theta}_{B'})\).
The corresponding energy differences in terms of the full interaction potential $V_I$ can be decomposed as \(\epsilon=\epsilon_+ +\epsilon_-\), accordingly.

The exact analytical result of Eq. \eqref{ST} leads to a complete classification of entanglement dynamics based on the symmetry of the interaction potential. For a symmetric coupling potential that \(V_I=V_{I+}\), the antisymmetric term \(\epsilon_- \) vanishes so that \(\epsilon=\epsilon_+\). Consequently, the expression of the linear entropy for both even and odd $t$ reduces to \(S_{\text{lin}}(t)=1-\left\langle \cos(t\epsilon)\right\rangle\). It reveals a characteristic timescale, \(t^*\sim1/\left\|\epsilon\right \|\), where \(\left\|\epsilon\right\|\equiv \sqrt{\left\langle \epsilon^2\right\rangle}\) is the root-mean-square of \(\epsilon\), quantifying the overall coupling strength.
Two distinct regimes thus emerge. At short times \(t < t^*\), a leading-order expansion of \(\cos(t\epsilon)\) predicts a quadratic entanglement growth,
\begin{equation}
S_{\text{lin}}(t)\approx \frac{\left\langle \epsilon^2\right\rangle}{2} t^2.
\label{Ssymmetric}
\end{equation}
Conversely, in the long-time limit, dephasing across different values of \(\epsilon\) causes the oscillatory term to average to zero, \(\left\langle \cos(t\epsilon)\right\rangle\to 0\).
This leads to the saturation of entanglement at a near-maximal value, \(S_{\text{lin}}\approx 1\).
The timescale \(t^*\) thus marks the crossover from initial quadratic increase to a saturated high-entanglement state.

A deeper insight into the origin of this quadratic growth can be gained by analyzing the purity, \(\mu_2(t) = 1-S_\text{lin}(t)\). It takes the form \(\mu_2(t)=\left\langle \cos(\epsilon t)\right\rangle \) and is dominated by the second moment at short times \(\mu_2(t)\approx1-\frac 12 \langle \epsilon^2\rangle t^2 \). This dephasing structure also underlies other fundamental quantum phenomena, such as the initial quadratic decay of the survival probability in many quantum processes~\cite{Sakurai2020}.

For an antisymmetric interaction, \(V_I=V_{I-}\), the symmetric term \(\epsilon_+\) vanishes and thus \(\epsilon = \epsilon_-\). Substituting this into Eq.~\eqref{ST} reveals a dramatically different behavior. At all even time steps the entropy is identically zero, $S_{\text{lin}}^{\text{even}}=0$, while at all odd time steps it takes a constant, non-zero value:
\begin{equation}
S_{\text{lin}}^{\text{odd}} = 1-\left\langle  \cos(\epsilon) \right\rangle.
\label{Slinodd}
\end{equation}
The entanglement therefore exhibits a striking, period-2 oscillatory behavior, alternating between
zero and a fixed value. This behavior is an entanglement analog of anti-resonance wavepacket dynamics of the single-body QKR. Its physical origin lies in the fact that the coupling-induced evolution operator responsible for entanglement generation becomes the identity operator every two steps (\(U_I^2=\mathbb I\)), such that \(U^2=U_\text{local}^2\). Since local evolution does not generate entanglement, this forces the system to oscillate between entangled and disentangled states.

For an asymmetric interaction potential, both \(\epsilon_+\) and \(\epsilon_-\) are nonzero. However, by expanding the exact result in Eq. \eqref{ST} to its leading order at short times, we have
\begin{equation}
\begin{aligned}
S_{\text{lin}}(t)
\approx \begin{cases}
\frac{\langle\epsilon _{+}^2\rangle}{2}t^2  , &t = 2m,\\
\frac{\langle\epsilon _{+}^2\rangle}{2}t^2+\langle\epsilon _{+}\epsilon _{-}\rangle t+\frac{\langle\epsilon _{-}^2\rangle}{2}, &t = 2m+1.\\
\end{cases}
\end{aligned}
\label{Slintapprox}
\end{equation}
It reveals a hybrid behavior: the entanglement entropy grows quadratically at even steps, controlled by the symmetric component through \(\langle \epsilon_+^2\rangle\), whereas at odd steps it acquires an additional constant offset set by \(\langle \epsilon_-^2\rangle\) and a linear time correction with coefficient set by the cross term \(\langle\epsilon _{+}\epsilon _{-}\rangle\). 

Indeed, this result for a generic coupling potential covers the two special cases discussed previously; i.e., by setting \(\epsilon_- =0 \) or \(\epsilon_+ =0 \), respectively, it reduces to the result for the symmetric or the antisymmetric potential. In particular, the constant offset \(\langle\epsilon _{-}^2\rangle/2\) for odd times is precisely the leading-order expansion of \(1-\left\langle \cos(\epsilon)\right\rangle\approx\langle  \epsilon _{-}^2\rangle/2 \) for the antisymmetric case.

\begin{table*}[t!]
\centering
\renewcommand{\arraystretch}{1.2}
    \begin{tabular*}{\textwidth}{@{\extracolsep{\fill}}lll}
        \hline
        \hline
        \textbf{}& \textbf{Symmetry condition} & \textbf{Dynamical regime}\\
        \hline
        \multirow{3}{*}{\textbf{Wavepacket spreading in \(p_j\)}}& Symmetric (\(V_j = V_{j+}\)) & Quadratic spreading: \(\sigma_j^2(t) \propto t^2\)\\
        &\cellcolor{gray!20}Antisymmetric (\(V_{j}=V_{j-}\)) &\cellcolor{gray!20}Oscillation: \(\sigma_j^2(t)\) oscillates with period 2\\
        & Asymmetric (mixed symmetry) &
        \makecell[l]{Hybrid: quadratic spreading modulated at odd-time steps}\\
        \hline
        \multirow{3}{*}{\textbf{Entanglement dynamics}} & Symmetric (\(V_I = V_{I+}\)) & Quadratic growth: \(S_{\text{lin}}(t) \propto  t^2\) before saturating\\
        &\cellcolor{gray!20}Antisymmetric (\(V_I = V_{I-}\))&\cellcolor{gray!20}Oscillation: \(S_{\text{lin}}(t)\) oscillates with period 2\\
        & Asymmetric (mixed symmetry) &
        \makecell[l]{Hybrid: quadratic growth modulated at odd-time steps\\
        before saturating}\\
        \hline
        \hline
    \end{tabular*}
\caption{Classification of dynamical regimes of many-body interacting kicked rotor systems in quantum resonances. The quantum resonance can be the principal, secondary, or a high-order resonance under the translational symmetry condition given by Eq.~\eqref{vtranssym} (see Sec. V). The wavepacket spreading in the momentum space of the $j$-th rotor, and the evolution of the entanglement entropy between two subsystems, are governed by the symmetries of the rotor's effective potential $V_j(\boldsymbol{\theta})$ and the inter-subsystem potential \(V_I(\boldsymbol{\theta})\), respectively.}
\label{tab:dynamics_classification}
\end{table*}

To summarize this section, we have identified a remarkable parallel between wavepacket and entanglement dynamics. Both are dictated by the symmetries of the relevant potentials (see Table I). This finding not only reveals the link between them, but also provides an alternative perspective for regulating many-body entanglement (e.g., utilizing antisymmetric interactions to realize periodic entangled states).

We emphasize that the three-regime characteristics identified in both wavepacket and entanglement dynamics are independent of the concrete form of the initial state. Rather, these characteristics arise qualitatively from the symmetric nature of corresponding potentials. The initial state merely  determines the values of the related parameters, such as  $\lambda_{j\pm}$ and  \(\langle  \epsilon _{\pm}^2\rangle\) governing the prefactors of quadratic growth and oscillation amplitudes.

Second, the wavepacket dynamics and entanglement dynamics in our model are independent, despite their apparent similarity. Indeed, in analyzing either of them, no reference to the other is required. Nevertheless, it is important to note that one can still infer one dynamics from the other to some extent, by exploiting the mechanism through which the potential symmetries act. 

For a rotor in either subsystem, say \(A\), its effective potential \(V_j(\boldsymbol{\theta})\) can be decomposed as 
\begin{equation}
    V_j(\boldsymbol{\theta}) = V_{j, A}(\boldsymbol{\theta}_A) + V_{j, I}(\boldsymbol{\theta}),
\end{equation}
where the first term $V_{j, A}(\boldsymbol{\theta}_A)$ denotes its local interacting potential within subsystem $A$, and the second term, $V_{j, I}(\boldsymbol{\theta})$ (assumed to be nonzero), denotes its interaction with the other subsystem. It is now clear that the inter-subsystem interaction $V_I(\boldsymbol{\theta})$ in fact imposes a ``selection rule'' on the symmetry of  $V_j(\boldsymbol{\theta})$ and thus on the dynamical regime of the wavepacket. To be specific, a symmetric $V_I(\boldsymbol{\theta})$ implies a symmetric $V_{j, I}(\boldsymbol{\theta})$, such that $V_j(\boldsymbol{\theta})$ cannot be antisymmetric, regardless of the symmetry of $V_{j, A}(\boldsymbol{\theta}_A)$. This implies that if the entanglement entropy grows quadratically, the wavepacket cannot exhibit period-2 oscillation in $p_j$. Similarly, an antisymmetric $V_I(\boldsymbol{\theta})$ rules out a symmetric $V_j(\boldsymbol{\theta})$, implying that if the entanglement entropy oscillates with period 2, the wavepacket cannot spread quadratically in $p_j$. For an asymmetric $V_I(\boldsymbol{\theta})$, the dynamical regime of the wavepacket then depends on the specific symmetry of $V_{j, I}(\boldsymbol{\theta})$. 

Conversely, because of such coupling in symmetries of $V_{j}$ and $V_{I}$, we can in some cases infer the regime of entanglement dynamics from that of wavepacket dynamics. For instance, if the wavepacket spreads quadratically in all momentum subspaces, i.e., $V_j(\boldsymbol{\theta})$ is symmetric for all rotors, then $V_I(\boldsymbol{\theta})$ must be symmetric, so that the entanglement entropy increases quadratically.  Similarly, if the wavepacket oscillates with period 2 in all momentum subspaces, then the entanglement entropy must oscillate with period 2 as well. 

Such mutual inferability, manifested to varying degrees, accounts for the intriguing symmetry-based  connection between the wavepacket and entanglement dynamics in our model.

\section{Illustrative examples}

In this section, we subject our analytical results summarized in Table I to numerical verification. To this end, we consider representative two-rotor models that have been extensively studied in the literature with a focus on the non-resonance regime. Here we focus on the unexplored resonance regime instead, in the hope to show the contrast effects of quantum resonance. First of all, for all various models and initial states having been simulated---most prominently the product of the rotors' coherent states, an initial state widely adopted in the literature---our theoretical predictions are verified with high precision.  Without loss of generality and for simplicity, we take the initial state $|\psi(0)\rangle=|p_1=0\rangle\otimes |p_2=0\rangle$ as our illustrative example throughout this work. This choice yields that $\alpha_{j\pm}=\kappa_j=0$ in all the studied models, so that $D_j(t)=0$ and $\sigma_{j}^2(t)=\langle p_{j}^2 (t)\rangle$. Accordingly, $\sigma_{1,2}^2(t)=\langle p_{1,2}^2 (t)\rangle$ correspond to the variances of the wavepacket, which are characterized solely by $\lambda_{j\pm}$. 

The first model we consider is the one in Refs. \cite{SP2020, Notar2018, Russomanno2023} with the potential
\begin{equation}
    V(\theta_1, \theta_2) = k_1\cos(\theta_1)+k_2\cos(\theta_2) + \xi\cos(\theta_1-\theta_2),
    \label{v12}
\end{equation}
where by definition, the effective potentials of the two rotors are $V_{1,2}=k_{1,2}\cos(\theta_{1,2}) + \xi\cos(\theta_1-\theta_2)$, respectively, and the coupling potential that entangles the two rotors is $V_I (\theta_1, \theta_2) = \xi\cos(\theta_1-\theta_2)$. 

When both rotors are set at the principal resonance with $\tau_{1,2}=4\pi$, by Eq. \eqref{thetaprime} the transformed angular coordinates are $\theta_{1,2}'=\theta_{1,2}$. As a result, both effective potentials $V_{1,2}$ and the coupling potential $V_I$ are symmetric under the transformation $U_\pi$. According to our theory, the wavepacket variance thus increases quadratically in the momentum spaces of both rotors and the entanglement entropy grows quadratically until time $t^*$. This is perfectly corroborated by our simulations (data presentation is omitted). Note that in Ref. \cite{SP2024}, only the case in which all rotors are placed at the principal resonance is investigated, the same as our setting here, and the quadratic growth of the linear entanglement entropy found for a similar two-rotor model therein is consistent with our analytical and numerical results.

\begin{figure}[t!]
    \centering
    \includegraphics[width=1\linewidth]{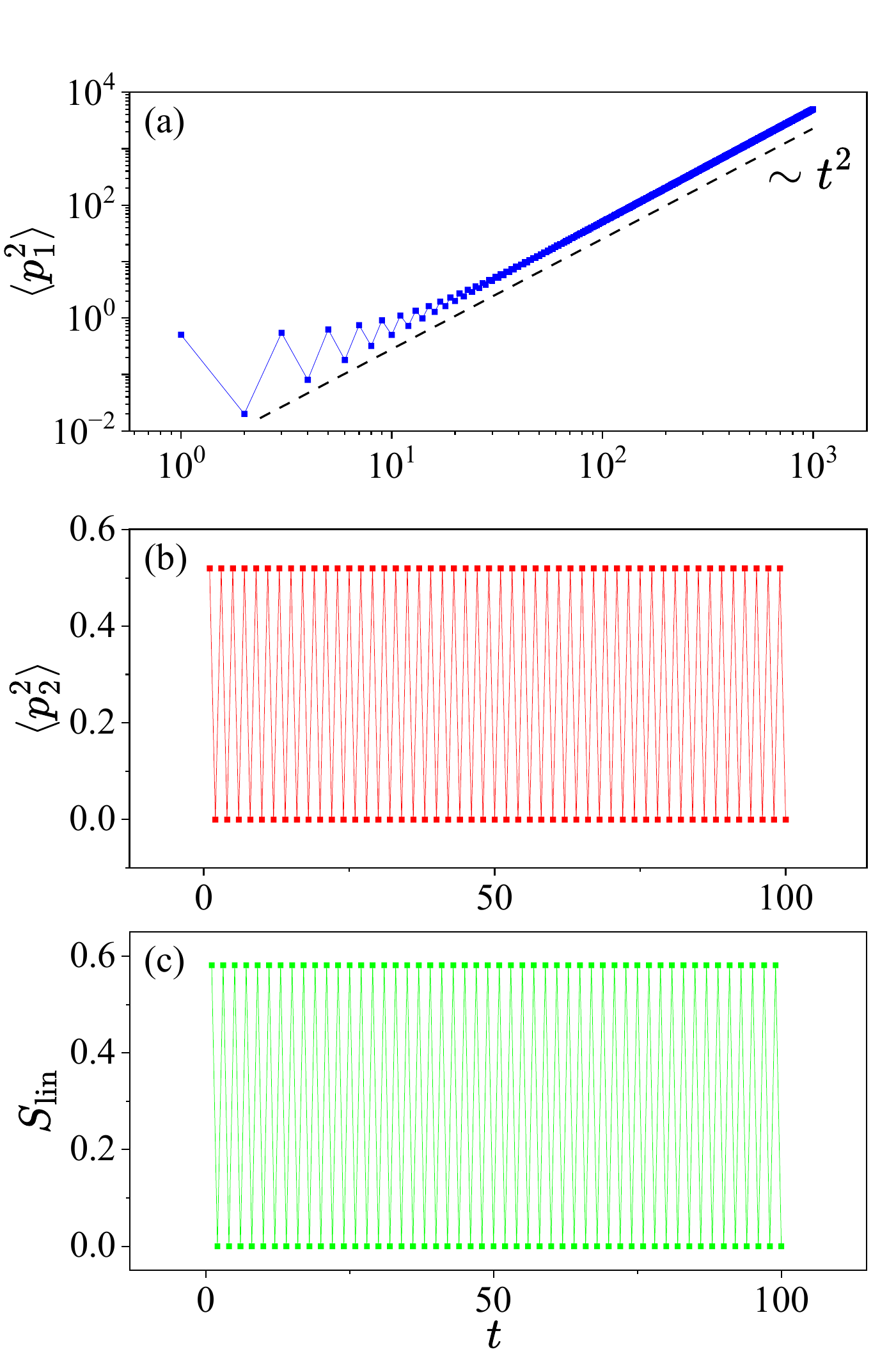}
\caption{Dynamics of the two-rotor system with potential \eqref{v12} and parameters $k_1=0.1$, $k_2=0.2$, and \(\xi=1\). The rotors are at the principal (\(\tau_1= 4\pi\)) and secondary (\(\tau_2= 2\pi\)) resonance, respectively. (a) The hybrid behavior of $\langle p_1^2 \rangle$ with ballistic growth of coefficient \(\lambda_{1+}=0.005\) and oscillations of magnitude \(\lambda_{1-} = 0.5\). (b) The oscillating behavior of $\langle p_2^2\rangle$ between zero and \(\lambda_{2-} = 0.52\). (c) The oscillating linear entanglement entropy between zero and $S_\text{lin}^{\text{odd}}= 1-\langle \cos(\epsilon)\rangle \approx 0.58$ [see Eq. \eqref{Slinodd}].}
  \label{fig:1abc}
\end{figure}

However, the phenomena are more striking when one rotor is at the principal resonance while the other is at the secondary resonance. To be concrete, let us put rotor 1 at the principal resonance (with $\tau_1 = 4\pi$) and rotor 2 at the secondary resonance (with $\tau_2 = 2\pi$), respectively, such that the transformed coordinates are $\theta_1'=\theta_1$ and $\theta_2' = \theta_2+\pi$. As a consequence, the effective potential for the first rotor, \(V_1 = k_1\cos(\theta_1) + \xi\cos(\theta_1-\theta_2)\), is neither symmetric nor antisymmetric under transformation $U_\pi$. Our theory thus predicts a hybrid wavepacket spreading behavior in the $p_1$ space, with the persistent quadratic growth of $\langle p_1^2 \rangle$ of coefficient \(\lambda_{1+}=k_1^2/2\) and the superposed oscillations set by \(\lambda_{1-}=\xi^2/2\). This prediction agrees perfectly with the simulation results shown in Fig.~\ref{fig:1abc}(a). Nevertheless, the effective potential for the second rotor, \(V_2 = k_2\cos(\theta_2) + \xi\cos(\theta_1-\theta_2)\), is purely antisymmetric under $U_\pi$, such that by our theory, the wavepacket is localized in the $p_2$ space that oscillates with an amplitude set by \(\lambda_{2-}=(k_2^2+\xi^2)/2\). Indeed, this oscillatory behavior is observed in simulations, as clearly illustrated in Fig.~\ref{fig:1abc}(b). As to the coupling potential \(V_I = \xi\cos(\theta_1-\theta_2)\), we can tell that it is also purely antisymmetric. This yields that \( \epsilon_+= 0\) and \(\epsilon = \epsilon_-\), implying a period-2 oscillation of the linear entanglement entropy, which is confirmed numerically as well [see Fig.~\ref{fig:1abc}(c)]. Incidentally, the case discussed here also serves as a good example to illustrate how the symmetries govern the physics underlying the rotors' effective potentials and coupling potential.

\begin{figure}
    \centering
    \includegraphics[width=1\linewidth]{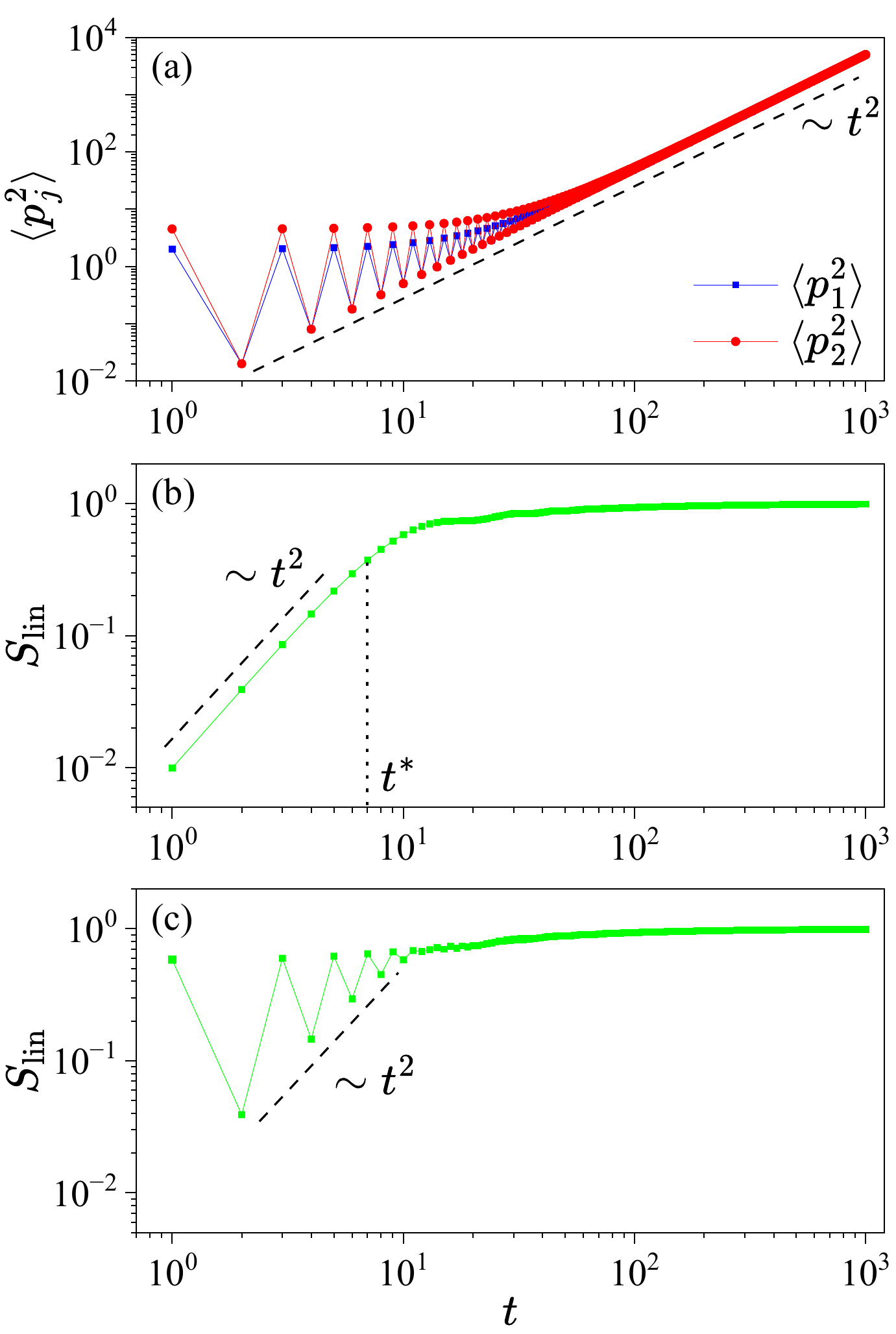}
    \caption{Dynamics of the two-rotor system with potential \eqref{v12} but parameters $k_1=2$, $k_2=3$, and \(\xi=0.1\). Both rotors are tuned at the secondary resonance (\(\tau_1=\tau_2= 2\pi\)). (a) The wavepacket exhibits hybrid spreading in both momentum spaces, where oscillations of amplitude $\lambda_{1-} =2$ and $\lambda_{2-} = 4.5$ are superimposed on the ballistic growth of coefficient $\lambda_{1+}=\lambda_{2+} = 0.005$. (b) The linear entanglement entropy exhibits a transition from quadratic growth to saturating. The analytically predicted crossover time \(t^*\) is marked by the vertical dotted line. (c) The linear entanglement entropy of a variant system with an additional potential term  \(\cos(2\theta_1-\theta_2)\) that shows the predicted superposition of quadratic growth and a period-2 oscillation.}
    \label{fig:2ab}
\end{figure}

\begin{figure}[t!]
    \centering
    \includegraphics[width=1\linewidth]{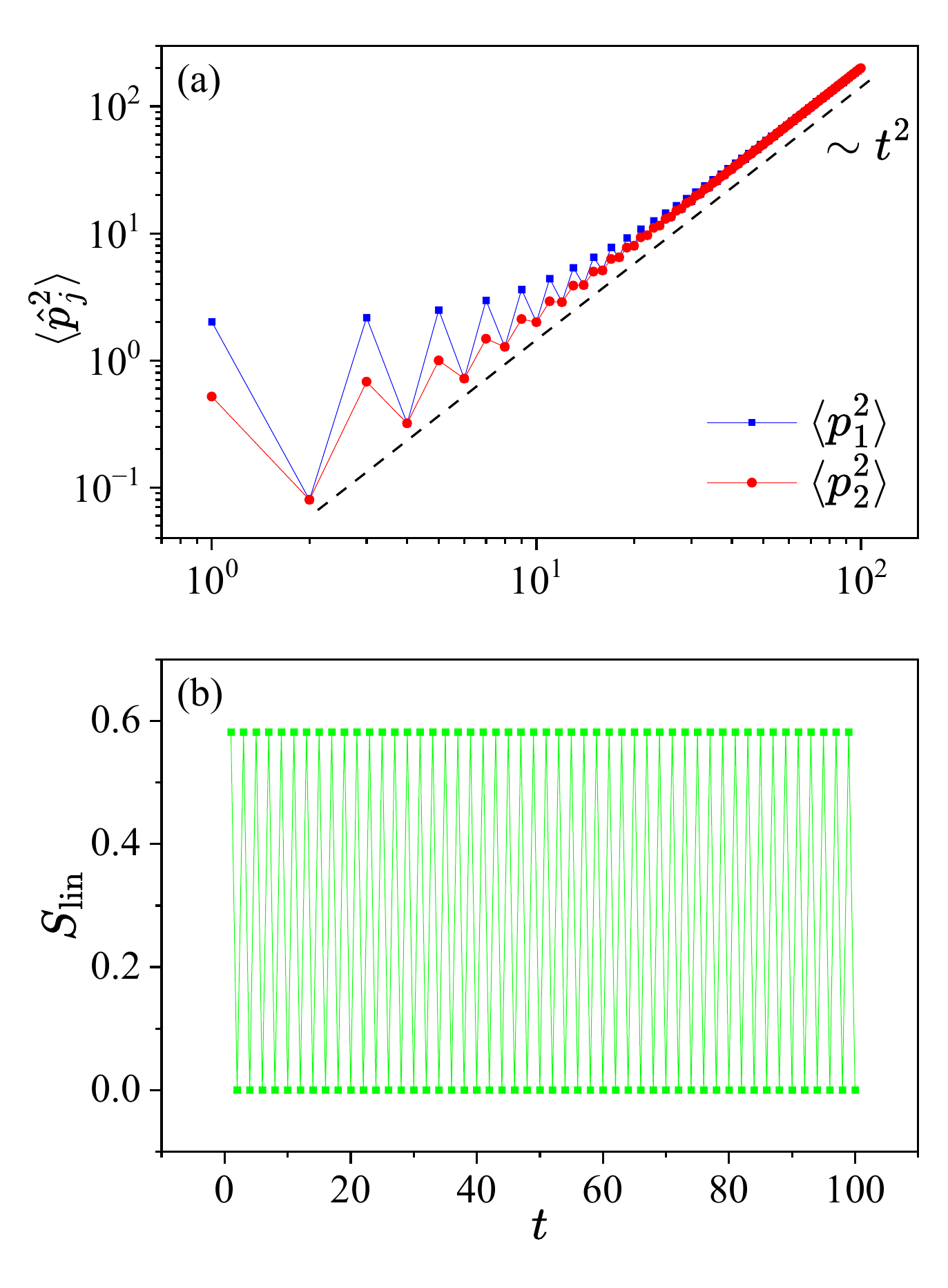}
    \caption{Dynamics of the extended, generalized model with potential \eqref{veg2} and parameters $k_1= k_2=0.1$ and \(\xi=1\). Both rotors are at the secondary resonance (\(\tau_1=\tau_2= 2\pi\)). (a) The wavepacket exhibits hybrid spreading in both momentum spaces, where for $\langle p_1^2 \rangle$ and $\langle p_2^2 \rangle$, oscillations of amplitude \(\lambda_{1-} =2\) and \(\lambda_{2-} = 0.5\) are superimposed respectively on the ballistic growth of coefficient \(\lambda_{1+}=\lambda_{2+} = 0.02\). (b) In contrast with the cases studied in Fig. 2, here the linear entanglement entropy oscillates between zero and \(S_\text{lin}^{\text{odd}}= 1-\langle \cos(\epsilon)\rangle\approx 0.58\) [see Eq. \eqref{Slinodd}] because of the antisymmetric interaction $V_I$.}
    \label{fig:3ab}
\end{figure}

Next, we turn to the case in which both rotors are at the secondary resonance with $\tau_1 =\tau_2= 2\pi$. It follows that $\theta_{1,2}'=\theta_{1,2}+\pi$ and the effective potentials for both rotors, \(V_1\) and \(V_2\), are neither symmetric nor antisymmetric under $U_\pi$. Consequently, we can predict that the wavepacket should exhibit a hybrid spreading behavior in both momentum spaces. For $\langle p_1^2 \rangle$, it should be characterized by coefficients \(\lambda_{1+}=\xi^2/2\) and \(\lambda_{1-}=k_{1}^2/2\), while for $\langle p_2^2 \rangle$, it should be characterized by \(\lambda_{2+}=\xi^2/2\) and \(\lambda_{2-}=k_{2}^2/2\). As confirmed in Fig.~\ref{fig:2ab}(a), the numerical results of $\langle p_1^2 \rangle$ and $\langle p_2^2 \rangle$ indeed exhibit a combination of quadratic growth and a periodic oscillation with the exact quantitative characteristics analytically suggested. Nevertheless, the behavior of the entanglement is markedly different, because the coupling potential \(V_I\) is purely symmetric instead. The entanglement is thus predicted to grow quadratically in an early stage -- i.e., \(S_{\text{lin}}(t) \approx \xi^2t^2\) [see Eq. \eqref{Ssymmetric}], considering that \(\langle \epsilon_+^2 \rangle = 2\xi^2\) -- followed by a transition to saturation at the crossover time \(t^*\sim1/(\sqrt 2\xi)\). As shown in Fig.~\ref{fig:2ab}(b), all these predicted features are well exhibited by the numerical results. In particular, the observed crossover time is in excellent agreement with our theoretical evaluation.

We have demonstrated all possible dynamical behaviors of the wavepacket and entanglement of the discussed two-rotor model. In particular, as $V_I$ can only be symmetric or antisymmetric, a hybrid entanglement dynamics is thus excluded. Therefore, in order to demonstrate the hybrid entanglement dynamics, we have to introduce a symmetry-breaking $V_I$. As an example, we modify the original $V_I$ by adding an extra term \(\cos(2\theta_1-\theta_2)\), while keeping all other settings the same as stated in the last paragraph. As expected, the resulting entanglement dynamics [shown in Fig.~\ref{fig:2ab}(c)] exhibits clearly the predicted superposition of quadratic growth and period-2 oscillation. Note that the modified $V_I$ does not change the asymmetric nature of $V_{1,2}$, hence the wavepacket spreading remains in the hybrid regime, with only stronger oscillatory components resulted from the additional term of $V_I$.

Note that all the analytical and numerical results presented so far have suggested that the symmetric nature of the relevant potentials, not their concrete forms, is crucial. To illustrate this point further, we consider a potential with high-frequency modes as an extended, generalized example:
\begin{equation}
    V(\theta_1, \theta_2) = k_1\cos(2\theta_1)+k_2\cos(2\theta_2)+\xi\cos(2\theta_1-\theta_2).
    \label{veg2}
\end{equation}
With this model, we avoid reiterating the dynamical behaviors already presented by the previous model but instead demonstrate a different, complementary set of behaviors. To this end, we tune both rotors to the secondary resonance (\(\tau_1 = \tau_2 = 2\pi\)) such that \(\theta_{1,2}'=\theta_{1,2}+\pi\). Under this $\pi$-translation, the coupling potential  \(V_I=\xi\cos(2\theta_1-\theta_2)\) is purely antisymmetric, while the effective potentials for the rotors ($V_1$ and $V_2$) are neither symmetric nor antisymmetric. Consequently, the wavepacket should exhibit a hybrid spreading behavior with $\lambda_{1+} = 2k_1^2$ and $\lambda_{1-} = 2\xi^2$ characterizing $\langle p_1^2 \rangle$, and $\lambda_{2+} = 2k_2^2$  and $\lambda_{2-} =\xi^2/2$ characterizing $\langle p_2^2 \rangle$. Meanwhile, for the entanglement, the linear entropy should undergo a period-2 oscillation between zero and \(1-\langle \cos(\epsilon)\rangle\). All these expectations are confirmed nicely again by numerical results shown in Fig.~\ref{fig:3ab}(a) and (b).

This striking contrast between wavepacket and entanglement dynamics is intriguing. While the wavepacket keeps spreading unboundedly, the entanglement's behavior is exactly periodic. Note that the oscillation amplitude \(1-\langle \cos(\epsilon)\rangle\) is determined by the parameters involved in $V(\theta_1, \theta_2)$, hence can be tailored. For an amplitude tuned close to one, the two rotors can be highly entangled from a purely disentangled state in just a single step, achieving an entanglement generating efficiency that can be arbitrarily close to its upper limit in principle. If the coupling between the two rotors is then switched off, they will remain in this highly entangled state, regardless of the ensuing wavepacket dynamics. The same interesting observation is that these two highly entangled rotors can be disentangled completely in a single step as well, resulting in an equally high disentanglement efficiency. On the other hand, if the amplitude is tuned close to zero, the two rotors are then effectively disentangled throughout. All these observations, including those in Fig. \ref{fig:1abc} and \ref{fig:2ab}, suggest clearly that  wavepacket and entanglement dynamics can be much more complicated and richer because of quantum resonance.

To summarize this section, we stress again that although our classifications for wavepacket and entanglement dynamics are seemingly ``parallel''---based on the symmetric nature of $V_j$ and $V_I$ (see Table I), respectively---they are connected in fact, because \(V_j\) comprises both the local potential and  coupling potential \(V_I\). Hence the symmetric nature of \(V_j\) also depends on \(V_I\) (see discussion at the end of Sec. III). For two-rotor models discussed here, \(V_j\) cannot be purely symmetric when \(V_I\) is antisymmetric, or purely antisymmetric when \(V_I\) is symmetric. This imposes a strong constraint on the allowed dynamics exhibited by the wavepacket and the entanglement. Particularly, if one rotor exhibits a period-2 oscillation, implying that its effective potential \(V_j\) is antisymmetric, then \(V_I\) must be antisymmetric as well. Consequently, the entanglement entropy must exhibit a period-2 oscillation as well. The case in which both $\langle p_1^2 \rangle$ and $\langle p_2^2 \rangle$ oscillate while the entanglement entropy does not can be equally ruled out.

\section{Generalization: high-order resonances}

In Sec. III, we discussed the wavepacket and entanglement dynamics when rotors are at the principal and secondary resonances. Here, we extend our analysis to include higher-order resonances. 

To this end, it is instructive to understand why the two lowest-order resonances are analytically amenable, a property that can be traced back to the single-body QKR.  In an $s$-order resonance single-body QKR with $\tau=4\pi r/s$, it has been shown that the complexity of its wavepacket dynamics originates from wavepacket splitting and proliferation \cite{Zou2024}. For instance, for an initial coherent-state wavepacket, after one iteration it splits into $\mathcal{N} (s)$ wavepackets along a line parallel to $\theta$-axis, with $\mathcal{N} (s)=s$ for odd $s$ or $\mathcal{N} (s)=s/2$ for even $s$. As a consequence, the number of the wavepackets proliferates exponentially, up to $\mathcal{N} (s)^t$ at most after $t$ steps. On the other hand, the proliferated wavepackets will interfere with each other because of the compactness of $\theta$-space, complicating the wavepacket dynamics further. For this reason, the high-order resonance wavepacket dynamics is analytically intractable even in the single-body case. For the two lowest-order resonances, however, $\mathcal{N}(1)=\mathcal{N}(2)=1$, implying that the wavepacket does not split and proliferate. In such a sense, these two cases are special and simple. 

Nevertheless, for a high-order resonance QKR, it is shown  that if the potential $V(\theta)$ has the symmetry that  $V(\theta)=V(\theta+\Delta \theta)$ with $\Delta \theta=2\pi/ \mathcal{N} (s)$, then, owning to the destructive interference induced by this symmetry, the proliferated wavepackets may cancel each other during their evolution, ensuring that the total number of wavepackets never exceeds $\mathcal{N} (s)$ \cite{Zou2024}.  The proliferation difficulty is thus avoided, and an analytical treatment becomes accessible again. 

Thus, our primary motivation here is to investigate whether one can analytically treat a many-body QKR at high-order resonances by introducing and exploiting such symmetries.  Suppose that it consists of $N$ rotors. For the $j$-th rotor, we set \(\tau_j=4\pi r_j / s_j\) with coprime integers \(r_j\) and \(s_j\). The rotors are then classified into two groups according to the parity of their resonance orders. If $e$ out of $N$ rotors have an even $s_j$, we use $\mathcal{E}$ to denote the set of their indices, \(\mathcal{E}\equiv \{j_1, j_2,\dots,j_e\}\). Namely, for $j\in\mathcal{E}$, $s_j$ is even. Note that \(\mathcal{E}\) and $e$ reduce to \(\mathcal{S}\) and $q$ defined previously when only principal- and secondary-resonance rotors are considered.

Mathematically, for a generic potential $V(\boldsymbol{\theta})$, the dynamics of the system is difficult to treat analytically, since the free evolution operator \(U_f=\exp(-i\sum_{j} {\tau_j p_j^2}/{2})\) and the kick operator \(U_k=\exp(-iV(\boldsymbol{\theta}))\) do not commute, in contrast to the previously studied case involving only the two lowest-order resonances. We therefore impose potential symmetries analogous to that in the single-body case to facilitate our analysis. Namely, for each angular coordinate $\theta_j$, $V(\boldsymbol{\theta})$ satisfies
\begin{equation}
V(\theta_1,\dots, \theta_j+\Delta\theta_j, \dots,\theta_N) =  V(\theta_1,\dots, \theta_j, \dots,\theta_N)
\label{vtranssym}
\end{equation}
with
\begin{equation}
\Delta\theta_j=
    \begin{cases}
     4\pi/s_j, &j\in \mathcal{E} ,\\
    2\pi/s_j, &\text{otherwise}.
    \end{cases}
    \label{deltathetai}
\end{equation}
Note that for a rotor at the principal or secondary resonance, $s_j = 1$ or $2$, we have $\Delta\theta_j = 2\pi$, hence this imposed symmetry condition is automatically satisfied owing to the periodicity $2\pi$ of angular coordinates. Therefore, it only applies in effect to rotors in higher-order resonances with \(s_j>2\).

To proceed, we employ the method developed in the study of the single-body in-resonance QKR~\cite{Zou2024}. First, we rewrite the evolution operator $U$ using the dressed version of $U_f$ and $U_k$, defined respectively as
\begin{equation}
U_{f}'\equiv U_{f}\exp(i\pi \sum_{j\in \mathcal{E}} p_j),~~~~~
U_{k}'\equiv \exp(-i\pi \sum_{j\in \mathcal{E}} p_j)U_{k}.
\end{equation}
Obviously, $U=U_f'U_k'$. A key observation is that, under the required symmetry condition specified in Eq.  \eqref{vtranssym}, the dressed operators commute, i.e.,  \(U_{f}'U_{k}'=U_{k}'U_{f}'\). Such a property is a natural many-body generalization of that for the single-body QKR \cite{Zou2024}, which can be shown straightforwardly. As a result, the operator for $t$-step evolution is given by
\begin{equation}
    U^t = (U_{f}')^t(U_k')^t,
\end{equation}
which factorizes into two operators that separately contain only free rotations or kicks, making further analytical investigation possible.

Partial simplification comes from \((U_f')^t\). It is diagonal in the momentum basis, hence its action does not change each rotor's kinetic energy. Meanwhile, it is free from the potential coupling the two subsystems, hence does not generate entanglement. Therefore, all non-trivial dynamics for both wavepacket and entanglement is governed by the second term, \((U_k')^t\), exclusively. More importantly, it is worth noting that \((U_k')^t\) has the same structure as the operator $U^t$ in the previously discussed case in which all rotors are at the two lowest-order resonances. In particular, it can be rewritten in the same form as Eq. \eqref{Ut}.

Because of this identical structure, the entire analytical framework developed in Sec. III can be directly translated to the present case. The resulting classifications of the dynamical regimes for both wavepacket and entanglement remain the same (see Tab. I). Therefore, the rich dynamical behaviors illustrated in the last two sections are universal even when rotors are placed in high-order resonances, provided the required symmetries in the potential are respected.

\begin{figure}
\centering
\vskip2.0mm
\includegraphics[width=1\linewidth]{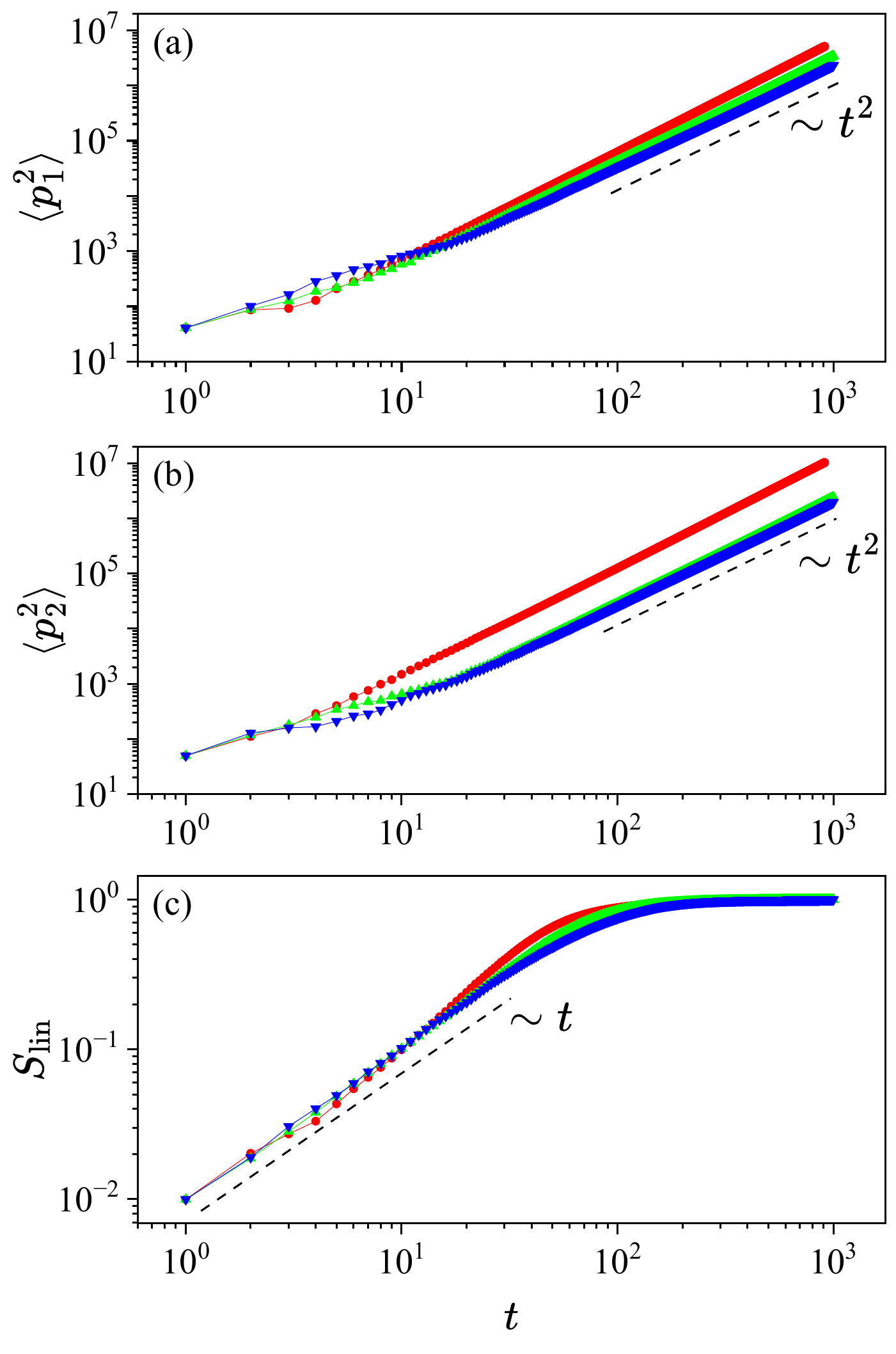}
\caption{Dynamics of three illustrative higher-order-resonance two-rotor models with $k_1=9$, $k_2=10$, and $\xi=0.1$. The three curves in each panel are for $\tau_1=4\pi/3$ and $\tau_2=4\pi/5$ (red),  $\tau_1=4\pi/13$ and $\tau_2=4\pi/15$ (green), and $\tau_1=4\pi/19$ and $\tau_2=4\pi/21$ (blue), respectively.}
\label{fig:HOR}
\end{figure}

Despite this solvable extension, it is interesting to explore what may occur in a general many-body model of higher-order resonance rotors in the absence of imposed potential symmetries. To this end, we perform extensive numerical simulations on the two-rotor model and find that the results are qualitatively consistent regardless of the resonance orders: the wavepacket variances grow quadratically, whereas the linear entanglement entropy grows linearly before saturating. 
In Fig. \ref{fig:HOR}, the simulation results for several typical examples are presented, where $V(\theta_1, \theta_2)$ given by Eq. \eqref{v12} is adopted. The quadratic growth of the wavepacket variances remains a consequence of quantum resonance, which endows the evolution operator with translational invariance. By contrast, the linear growth of entanglement entropy can be attributed to the complex wavepacket dynamics arising from wavepacket proliferation and interference, as revealed by comparison with the analytically solvable cases discussed previously. 

\section{Robustness analysis}

The dynamical classifications for wavepacket and entanglement have been established based on the quantum resonance condition, i.e., for each rotor, \(\tau_j\) is a precise rational multiple of \(4\pi\). A physically relevant question is how robust the predicted dynamical behaviors are when the quantum resonance condition is not fulfilled exactly. To answer this question, in this section we study how the wavepacket and entanglement dynamics would respond to deviations from the resonance condition. We set \(\tau_j=\tau_j^0+\delta\tau_j\), where \(\tau_j^0=4\pi r_j/s_j\) and  \(\delta\tau_j\) is incommensurate with \(4\pi\), by taking the offset \(\delta\tau_j\) as a deliberate perturbation to probe the robustness of the predicted dynamical behaviors. All our extensive numerical studies suggest that for small but non-zero perturbations, the system's wavepacket and entanglement dynamics are almost indistinguishable initially from their counterparts in exact resonance and the agreement time increases as the perturbations decrease, signifying their robustness consistently.

\begin{figure}[t!]
    \centering
    \includegraphics[width=1\linewidth]{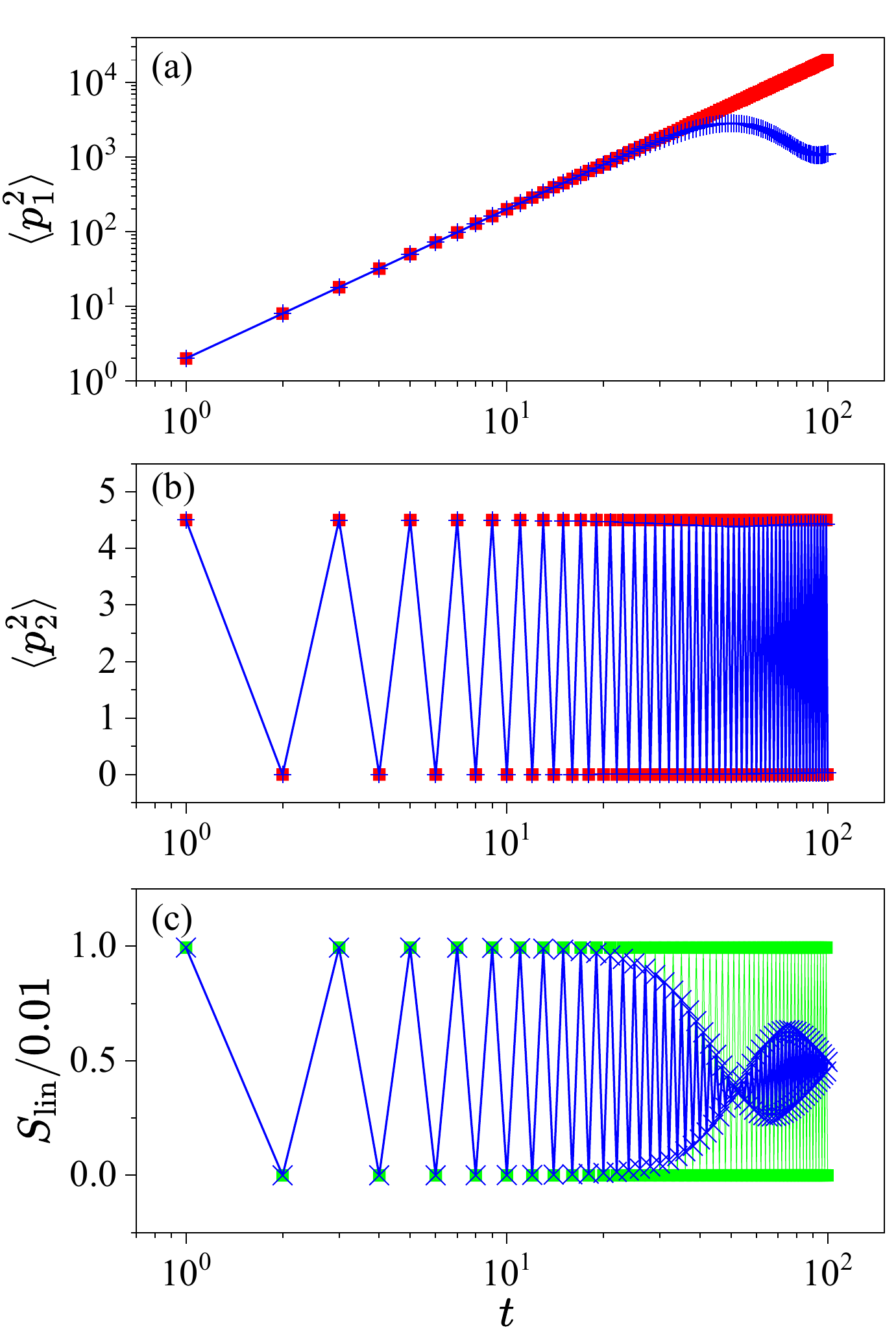}
    \caption{The simulation results for the two-rotor model (see text). In (a) and (b), time dependence of $\langle p_1^2\rangle$ and $\langle p_2^2\rangle$ are shown, respectively, where the results for the ideal resonance case (red solid squares) are compared against the perturbed case (blue pluses). (c) The same as (a) and (b) but for the linear entanglement entropy, where  green solid squares and blue crosses are for the ideal resonance and the perturbed case, respectively. In the simulation, the detuning strengths are \(\delta \tau_1 = \delta\tau_2 = 10^{-3}\). Other parameters are \(k_1=2\), \(k_2 = 3\), and \(\xi = 0.1\). For the undetuned entanglement entropy, the oscillating amplitude \(S_\text{lin}^{\text{odd}}= 1-\langle \cos(\epsilon)\rangle\approx 0.0099\) [see Eq. \eqref{Slinodd}]. }
    \label{fig:detuneab}
\end{figure}

As a concrete and typical example, we invoke the two-rotor model with the potential given in Eq.~\eqref{v12}, where the two rotors are at the principal and secondary resonance, respectively. Here the rotors are detuned slightly with $\tau_1= 4\pi + \delta\tau_1$ and $\tau_2=2\pi+\delta\tau_2$. The simulation results are presented in Fig.~\ref{fig:detuneab}, where we can see that all $\langle p_1^2 \rangle$, $\langle p_2^2 \rangle$, and the linear entropy follow their resonance counterparts fidelitously up to about time $t=20$ without any significant deviations, then the deviations set in and grow as expected.

\begin{figure}
    \centering
    \vskip2.0mm
    \includegraphics[width=1\linewidth]{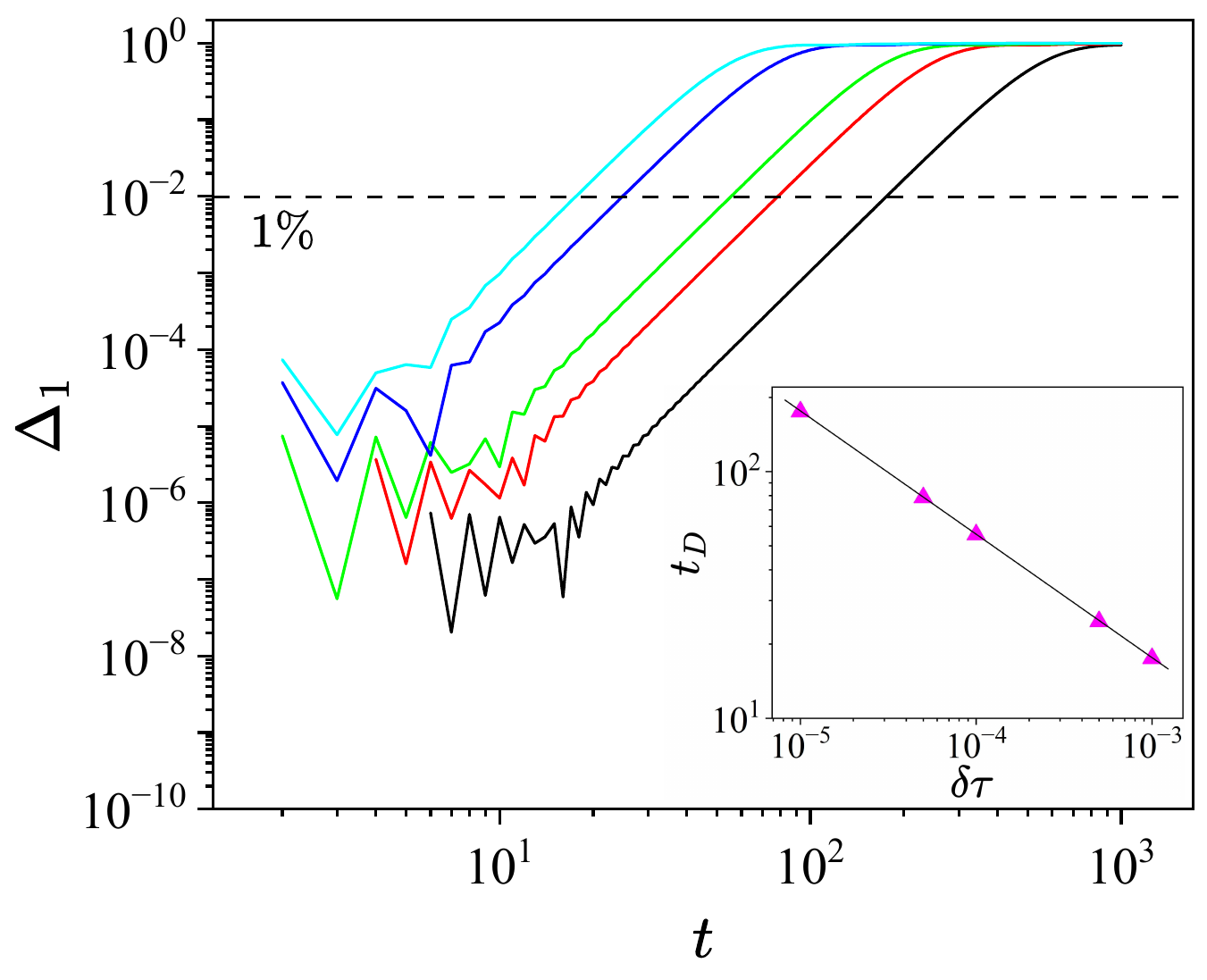}
    \caption{From top to bottom, the curves in the main panel are for the relative deviations of $\langle p_1^2\rangle$ at various detuning strengths for the two-rotor model, $\delta\tau = 10^{-3}$, $5\times 10^{-4}$, $10^{-4}$, $5\times10^{-5}$, and $10^{-5}$, respectively, where $\delta\tau_1 =\delta\tau_2=\delta \tau$. The inset shows the agreement time \(t_D\) against the detuning \(\delta\tau\) in logarithmic scales, where the data points fall on a straight line of slope \(-1/2\). Other parameters are the same as in Fig. \ref{fig:detuneab}.}
    \label{fig:relativediff}
\end{figure}

To scrutinize how the deviations develop in time and depend on perturbations, without loss of generality, we take the relative deviation of the kinetic energy of rotor 1 as a deviation measure,
\begin{equation}
    \Delta_1
    = \frac{\left | \langle {p}_1^2\rangle_{0}-\langle {p}_1^2\rangle\right|}{\langle {p}_1^2\rangle_0},
\end{equation}
where \(\langle {p}_1^2\rangle_{0}\) is the kinetic energy of rotor 1 in the corresponding undetuned system in quantum resonance. Still taking the model studied in Fig.~\ref{fig:detuneab} as an example, we show in Fig.~\ref{fig:relativediff} the deviation measure as a function of time at various detunings. Indeed, in general the stronger the detuning strength, the larger the deviation. In particular, over a time range that expands as the detunings decrease, the deviation measure grows in time as $\sim t^4$.

The results in Fig.~\ref{fig:relativediff} suggest that, for given detunings, the deviation can be neglected in a sufficiently short time, such that the dynamics of the detuned system can be regarded as agreeing with its undetuned, in-resonance counterpart. As a working definition of this agreement time, denoted \(t_D\), we set it to be the time when the deviation measure reaches a given threshold. In the inset of Fig.~\ref{fig:relativediff}, the dependence on detunings of \(t_D\) measured with a low threshold of  \(1\%\) is presented, where the scaling \(t_D\sim (\delta\tau)^{-1/2}\) can be clearly recognized. Note that this scaling applies equally to other deviation measures, e.g., the one defined in terms of $\langle p_2^2\rangle$ or entanglement entropy $S_{\text{lin}}$. Also note that, although we have fixed \(\delta\tau_1 = \delta\tau_2=\delta\tau\) here, for other more general detunings such that \(c_1 \delta\tau_1 = c_2 \delta\tau_2=\delta\tau\) with \(c_1\) and \(c_2\) being two given constants of the same order, the scaling \(t_D\sim (\delta\tau)^{-1/2}\) applies as well.

Based on these numerical analyses, it can be concluded that the analytical predictions for the in-resonance many-body kicked rotor models are robust against weak destruction of the resonance condition, in the sense that the time scale for which they remain valid diverges as the detuning strength tends to vanish. In fact, for other dynamical regimes of the example model and other models we have extensively checked with simulations, this conclusion holds throughout. As a consequence, it can be conjectured that the peculiar dynamical characteristics owing to resonance should be accessible conditionally in practice. In particular, when the entanglement entropy oscillates with period 2, the rapid entanglement and disentanglement occur in just two steps, allowing for a far less stringent detuning strength for observation and state preparation. Our theoretical results hence may have implications for future experimental study and applications.

\section{Many-body kicked tops in quantum resonance}

The conventional single-body quantum kicked top (QKT) \cite{Haake1987} is another paradigmatic model in quantum chaos. It shares essentially the same physical properties as the QKR: as the kicking strength increases, their classical dynamics shift from the integrable regime, to the integrable-chaotic mixed regime, and then to the fully chaotic regime. Furthermore, the two systems are even closely related in form, such that the QKR emerges as a limiting case of the QKT \cite{Haake1988}.  The only remarkable difference lies in their phase space geometries: a cylinder for the QKR versus a sphere for the QKT. Because of the compactness of the latter, the Hilbert space of the QKT is finite, making the QKT more feasible for numerical studies.  Nevertheless, it should be noted that the compactness of the sphere imposes a time bound for observing wavepacket spreading in the QKT (see below), whereas that in the QKR (in the momentum direction) could be unbounded. 

Recently, quantum resonance was also identified in the single-body QKT \cite{Zou2022}, providing further evidence of the essential similarity between the two systems. Taking advantage of this progress, in this section, we attempt to extend quantum resonance to many-body kicked top models and investigate the resulting wavepacket and entanglement dynamics. Our motivations are two-fold: one is to understand the generality of our findings for the in-resonance many-body QKR, and the other is to open a research direction for the QKT studies, in view of its profound implications for the related frontier topics in quantum science. The main conclusion is that the three-regime dynamics identified for wavepacket and entanglement in the rotor model can be directly translated to the top model, except for  the only minor time restriction imposed on wavepacket spreading arising from the compactness of the latter's phase space.

Suppose there are \(N\) tops in the model, each with a total angular momentum quantum number \(j_{\text{tot}}\), such that the sub-Hilbert space for each top has a dimension \(d = 2j_{\text{tot}}+1\). As before, we adopt the dimensionless units where $\hbar=T=1$ ($T$ is the period of the kicks) and omit them from the related expressions. The one-step evolution of the system is governed by the operator \(U = U_f U_k\), with \(U_k\) being responsible for the kick and \(U_f\) for the motion between two consecutive kicks. The latter is generated by the Hamiltonian \(H_f(J_{1x}, \dots, J_{Nx})\), which is assumed to depend on the \(x\)-components of the angular momenta. For brevity, we denote \(\mathbf{J}_x \equiv (J_{1x}, \dots, J_{Nx})\), such that \(H_f(J_{1x}, \dots, J_{Nx})=H_f(\mathbf{J}_x)\). Note that in the conventional QKT, \(H_f(\mathbf{J}_x)\)  represents the free rotation of the top; here we assume a general physically allowed form of \(H_f(\mathbf{J}_x)\).

The kick operator is a nonlinear twist. The same as in the single-body QKT, we assume that it depends only on the \(z\)-components of the angular momenta,
\begin{equation}
U_k = \exp\left(-i \sum_{n=1}^N \frac{\beta_n}{2j_{\text{tot}}} {J}_{nz}^2\right),
\end{equation}
where \(\beta_n\) controls the kick strength on the $n$-th top. Analogous to the rotor system, quantum resonance condition for the $n$-th top is that \(\beta_n\) is a rational multiple of $4\pi$~\cite{Zou2022}. In the following we restrict ourselves to focus on the two lowest-order resonances, as generalization to incorporate higher resonances is also parallel to that in the rotor model.

Because of the integer spectrum of \({J}_{nz}\) (since $\hbar$=1), for principal-resonance tops with \(\beta_n = 4\pi j_{\text{tot}}\), their contribution to the kick operator \(U_k\) reduces to the identity. Hence effective contributions only come from the secondary-resonance tops with \(\beta_n = 2\pi j_{\text{tot}}\), such that
\begin{equation}
U_k = \exp\left(-i \pi \sum_{n \in \mathcal{S}} J_{nz}\right),
\end{equation}
where \(\mathcal{S}\) denotes the set of indices of secondary-resonance tops. This kick operator generates a rotation by \(\pi\) around its \(z\)-axis for each top in \(\mathcal{S}\), leading to the commutation relation
\begin{equation}
U_k \exp(-i H_f(\mathbf{J}_x)) = \exp(-i H_f(\mathbf{J}_x')) U_k,
\end{equation}
where \(\mathbf{J}_x'\equiv(J'_{1x}, \dots, J'_{Nx})\) is given by
\begin{equation}
J_{nx}' = \begin{cases}
- J_{nx}, & n \in \mathcal{S}, \\
J_{nx}, & \text{otherwise}.
\label{jprime}
\end{cases}
\end{equation}
The evolution over two steps then takes a similar structure as in Eq.~\eqref{u2rotor}, given by:
\begin{equation}
    \begin{aligned}
        U^2 &= U_f U_k U_f U_k \\
    &= \exp\left(-i ( H_f(\mathbf{J}_x) + H_f(\mathbf{J}_x') ) \right).
    \end{aligned}
\end{equation}
Obviously, as $J_{nx}$ commutes with $U^2$, it is conserved throughout, hence the wavepacket only spreads in other two directions of a top's angular momentum space.  An analogous analysis as performed in the rotor model allows us to classify the wavepacket dynamics based on the symmetry of \( H_f\) with respect to the transformation defined in Eq.~\eqref{jprime}. To be specific, taking $J_{nz}$ as an illustration and denoting by $H_{nf}(\mathbf{J}_x)$ the effective Hamiltonian of the $n$th top, which consists of only the  terms in $H_{f}(\mathbf{J}_x)$ that involve ${J}_{nx}$. It can be decomposed as $H_{nf}(\mathbf{J}_x)=H_{nf+}(\mathbf{J}_x)+H_{nf-}(\mathbf{J}_x)$, with the symmetric and antisymmetric components to be \(H_{nf+}(\mathbf{J}_x)\equiv [H_{nf}(\mathbf{J}_x)+H_{nf}(\mathbf{J}_x')]/2\) and \(H_{nf-}(\mathbf{J}_x)\equiv [H_{nf}(\mathbf{J}_x)-H_{nf}(\mathbf{J}_x')]/2\), respectively. The \(t\)-step evolution operator can be written as \(U^t=\exp(-i\mathcal{H}_{tnf})\) for even $t$ and  \(U^t=U_k\exp(-i\mathcal{H}_{tnf})\) for odd $t$ with the effective Hamiltonian
\begin{equation}
    \begin{aligned}
\mathcal{H}_{tnf} (\mathbf{J}_x) & = \begin{cases}
tH_{nf+} (\mathbf{J}_x) &t = 2m,\\
tH_{nf+} (\mathbf{J}_x)-H_{nf-} (\mathbf{J}_x), &t = 2m+1.
\end{cases}
\end{aligned}
\end{equation}

To capture the wavepacket dynamics, we adopt the mean and the mean squared displacement of the wavepacket, i.e., 
\begin{equation}
\begin{aligned}
    D_{nz}(t) &\equiv \langle \psi(0) | J_{nz}(t) - J_{nz}(0) |\psi(0) \rangle \\
    &= \langle \psi(0) | (U^t)^\dagger J_{nz}(0) U^t - J_{nz}(0) | \psi(0) \rangle,
\end{aligned}
\end{equation}
and 
\begin{equation}
\begin{aligned}
    \sigma_{nz}^2(t) &\equiv \langle \psi(0) | [J_{nz}(t) - J_{nz}(0)]^2 |\psi(0) \rangle\\
    &= \langle \psi(0) | [(U^t)^\dagger J_{nz}(0) U^t - J_{nz}(0)]^2 | \psi(0) \rangle,
\end{aligned}
\end{equation}
where $| \psi(0) \rangle$ is the initial state.  They are in turn characterized by the following parameters:
\begin{equation}
\alpha_{n\pm}^{(z)} \equiv \langle \psi(0) | \mathcal{D}_{n\pm}^{(z)} | \psi(0) \rangle,
\end{equation}
\begin{equation}
\lambda_{n\pm}^{(z)} \equiv \langle \psi(0) | [\mathcal{D}_{n\pm}^{(z)}]^2 | \psi(0) \rangle,
\end{equation}
and
\begin{equation}
\kappa_n^{(z)} \equiv \frac{1}{2} \langle \psi(0) | \{ \mathcal{D}_{n+}^{(z)}, \mathcal{D}_{n-}^{(z)} \} | \psi(0) \rangle ,
\label{kappa}
\end{equation}
where \(\mathcal{D}_{n\pm}^{(z)} = -i [J_{nz}, H_{nf\pm}(\mathbf{J}_x)] \) represents the effective displacement operator, and  $\{\cdot, \cdot\}$ in Eq. \eqref{kappa} denotes the anticommutator.
If \(H_{nf}\) is purely symmetric that \(H_{nf} = H_{nf+}\), we have \(\alpha_{n-}^{(z)} = \lambda_{n-}^{(z)}  = \kappa_n^{(z)} = 0\); the wavepacket drifts linearly in time with the mean displacement growing as \(D_{nz}(t) \approx t \alpha_{n+}^{(z)} \). Meanwhile, the mean squared displacement grows as 
\(\sigma_{nz}^2(t) \approx t^2 \lambda_{n+}^{(z)}\). If instead \(H_{nf}\) is purely antisymmetric, i.e., 
\(H_{nf} = H_{nf-}\),  we then have  that \(\alpha_{n+}^{(z)} = \lambda_{n+}^{(z)}  = \kappa_n^{(z)} = 0\), so that  both the mean and mean squared displacements oscillate between zero and a constant as
\begin{equation}
D_{nz}(t) \approx 
\begin{cases}
0, & t = 2m, \\ -\alpha_{n-}^{(z)}, & t = 2m+1,
\end{cases} 
\end{equation} 
\begin{equation}
\sigma_{nz}^2(t) \approx \begin{cases} 0, & t = 2m, \\ \lambda_{n-}^{(z)}, & t = 2m+1.
\end{cases} 
\end{equation}
For an asymmetric \(H_{nf}\), the mean displacement is dominated by the linear growth driven by \(H_{nf+}\), modulated by a constant offset at odd times owing to \(H_{nf-}\),
\begin{equation}
D_{nz}(t) \approx \begin{cases}
t \alpha_{n+}^{(z)}, & t = 2m, \\
t\alpha_{n+}^{(z)} - \alpha_{n-}^{(z)}, & t = 2m+1.
\end{cases}
\end{equation}
Similarly, 
\begin{equation}
\sigma_{nz}^2(t) \approx 
\begin{cases}
t^2 \lambda_{n+}^{(z)}, & t = 2m, \\
t^2 \lambda_{n+}^{(z)} - 2t \kappa_n^{(z)} + \lambda_{n-}^{(z)}, & t = 2m+1.
\end{cases}
\end{equation}

Note that to derive these results, we have invoked the linearized operator evolution
\begin{equation}
    (U^t)^\dagger J_{nz} U^t \approx J_{nz}-i [J_{nz}, \mathcal{H}_{tnf}]
\end{equation}
by retaining only the first-order term in the expansion of $(U^t)^\dagger J_{nz} U^t$. This is a good approximation when the wavepacket is close to the equator ($J_{nz}=0$) of the spherical phase space of the $n$-th top.  Therefore, the analytical predictions given here are for an initial state such that $\langle \psi(0)| J_{nz}| \psi(0)\rangle\approx 0$ and $\langle \psi(0)| J_{nz}^2| \psi(0)\rangle\approx 0$. Also note that a wavepacket cannot spread indefinitely. For the initial state considered here, the wavepacket will slow its expansion upon reaching the poles of the spherical phase space at a time $t_s$, which can be estimated from the condition $\sigma^2_{nz}(t_s)\sim j_{\text{tot}}^2$. Another constraint imposed by the compactness of the spherical phase space manifests itself in the long-time limit, known as quantum revival: the dynamics of the system becomes exactly periodic when the Hamiltonian contains terms up to quadratic order \cite{Parker1986, Robinett2004}. 

Now let us turn to discuss the entanglement dynamics. We partition the system into two subsystems, \(A\) and \(B\), and denote the corresponding \(x\)-component angular momentum vectors as \(\mathbf{J}_{xA}\) and \(\mathbf{J}_{xB}\), respectively. Thus \(H_f\) is decomposed into local and interacting parts, \(H_f(\mathbf{J}_x) = H_{f,A}(\mathbf{J}_{xA}) + H_{f,B}(\mathbf{J}_{xB}) + H_{f,I}(\mathbf{J}_{x})\), with the entanglement dynamics governed exclusively by the interaction \( H_{f,I}\). If \(H_{f,I}\) is symmetric with respect to the transformation of Eq. \eqref{jprime}, 
i.e., \( H_{f,I}(\mathbf{J}_x')=H_{f,I}(\mathbf{J}_x)\), then the effective evolution operator governing the entanglement dynamics is
\begin{equation}
    U_I = \exp(-iH_{f,I}(\mathbf{J}_x)).
\end{equation}
Note that similar to the rotor model, the eigenstates of this operator have a tensor-product form
\begin{equation}
    |m_{1x},\dots,m_{Nx}\rangle\equiv|m_{1x}\rangle\otimes\dots\otimes|m_{Nx}\rangle,
\end{equation}
where \(|m_{nx}\rangle\) represents an eigenstate of \(J_{nx}\) for the \(n\)-th top, i.e., \(J_{nx}|m_{nx}\rangle = m_{nx}|m_{nx}\rangle\).
This tensor-product structure of the eigenstate leads to a quadratic growth of the linear entropy in its early developing stage (see Appendix B). Conversely, if \(H_{f,I}\) is antisymmetric, i.e., \( H_{f,I}(\mathbf{J}_x')=-H_{f,I}(\mathbf{J}_x)\), the effective evolution operator over two steps becomes the identity, \(U_I^2 =\mathbb I\), leading to a period-2 oscillating entanglement dynamics. Similar to the kicked rotor case, if \(H_{f,I}\) is neither symmetric nor antisymmetric, a hybrid entanglement dynamics takes over.

\begin{figure}
\centering
\vskip2.0mm
\includegraphics[width=1\linewidth]{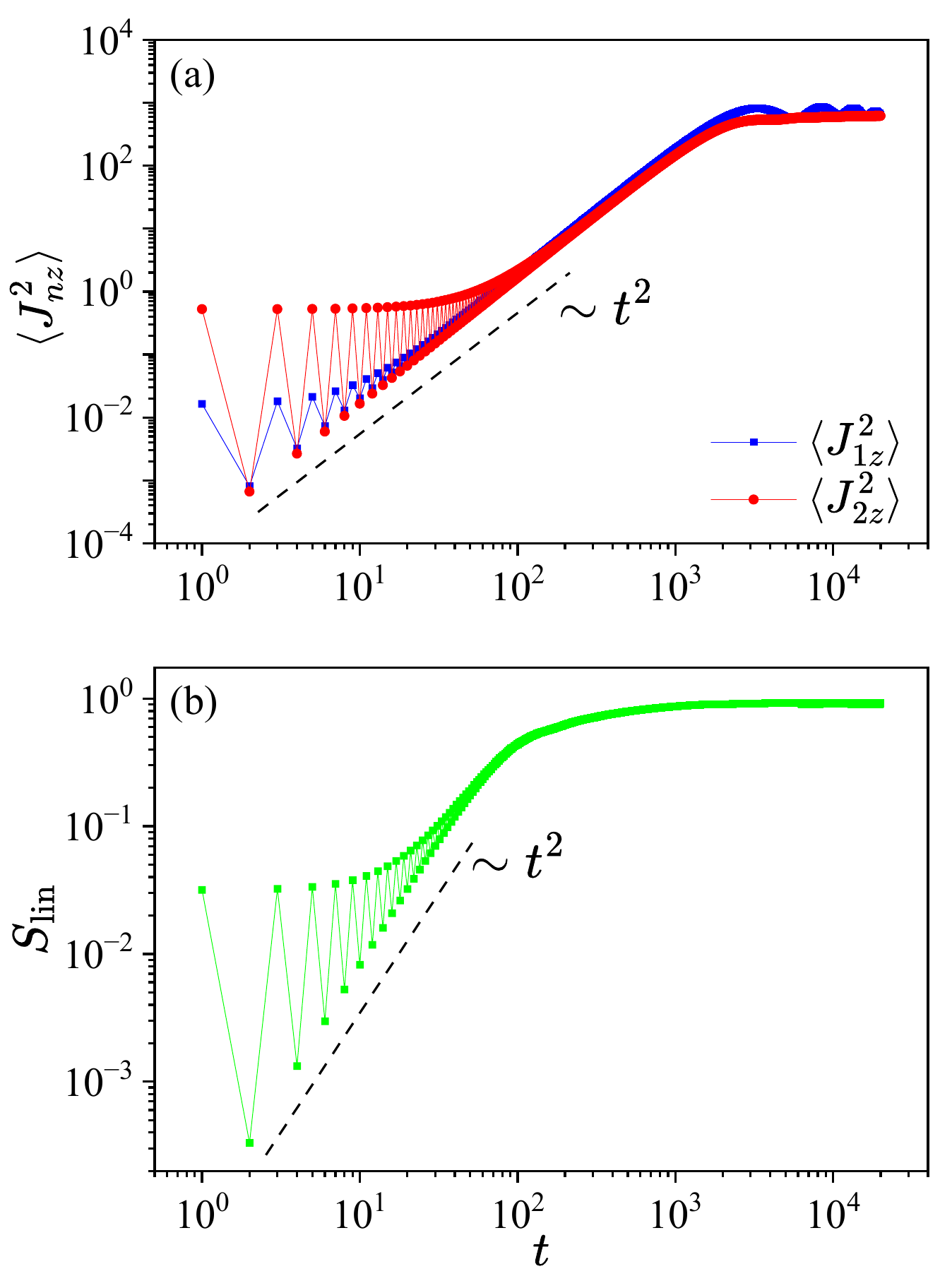}
\caption{Dynamics of the two-top model with $\alpha_1=0.0001$, $\alpha_2=0.02$, $\alpha_{12}=0.005$, $\alpha_{122}=0.0005$ [see Eq. \eqref{Hf}], and $j_\text{tot}$=50. Top 1 and 2 are in principal and secondary resonance, respectively.}
\label{fig:KT}
\end{figure}

As an example, in Fig. \ref{fig:KT} we present the simulation results for a two-top model with
\begin{equation}
    H_f = \alpha_1 J_{1x}+\alpha_2 J_{2x} 
    +\frac{ \alpha_{12}} {j_\text{tot}} J_{1x}J_{2x}
    +\frac{\alpha_{122}} {j_\text{tot}^2} J_{1x}J_{2x}^2.
\label{Hf}
\end{equation}
We impose the principal and secondary resonance conditions on the two tops by setting $\beta_1=4\pi j_\text{tot}$ and $\beta_2=2\pi j_\text{tot}$, and adopt $|J_{1z}=0\rangle\otimes |J_{2z}=0\rangle$ as the initial state, such that $D_{nz}(t)=0$ and $\sigma_{nz}^2(t)=\langle J_{nz}^2\rangle$. As all the effective potentials of the two tops and their interaction potential are asymmetric, we predict that $\langle J_{1z}^2\rangle$, $\langle J_{2z}^2\rangle$ should exhibit hybrid behavior, until they reach the bound $\sim  j_\text{tot}^2$ owing to the compactness of the phase space. Meanwhile, the entanglement dynamics should also lie in the hybrid regime. All these predictions are soundly corroborated.

\section{Summary and outlook}

In summary, we have perfomed a thorough and systematical investigation of a generic many-body kicked rotor system under quantum resonance. The predicted three-regime characteristic dynamical behaviors for both wavepacket and entanglement are well confirmed by simulations. Notably, the symmetry-based conection between the wavepacket and entanglement dynamics is elucidated.  

The revealed interesting entanglement dynamics, including ballistic growth, oscillation with period 2, and their superposition, are novel and enrich the library of known entanglement dynamics. It suggests that the entanglement dynamics could be much richer, and these additional entanglement dynamics may provide a different perspective for probing the profound aspect of quantum entanglement. It is also interesting to examine the known entanglement properties with our models, such as entanglement swapping, or to explore new possibilities, such as entanglement control, by, e.g., switching on and off certain tailored kicking potential terms during evolution.

The in-resonance many-body kicked rotor model and its parallelly constructed kicked top counterpart introduced in this work provide two analytically solvable nonequilibrium models for quantum many-body physics. Given that analytically solvable many-body models are rare and valuable, they are expected to serve as useful platforms for future studies and offer new insights into quantum many-body science. In this regard, previous studies based on single-body models, especially those on the single-body kicked rotor and kicked top and their variants, can be extended immediately to the many-body regime. In this context,  frontier topics of particular interest include, e.g., out-of-time-ordered correlation \cite{Galitski2017}, quantum fidelity \cite{Raizen2005}, non-Hermitian dynamics \cite{Longhi2017, Zhao2019}, and quantum-classical correspondence in deep quantum regime \cite{Zou2024, Huang2025}.

Finally, a promising many-body kicked rotor model for the experimental verification and further extension of our work is the kicked Lieb–Liniger rotor system, where the rotors are coupled via contact ($\delta$-function) interactions. Recently, an experimental study has demonstrated that under generic off-resonance conditions, this system enters a many-body localization phase \cite{Ying2025}. However, for the two-rotor version of this system \cite{Flach2017}, our study shows that when the two rotors are set to the principal or secondary quantum resonance, both the wavepacket dynamics in the center-of-mass momentum and relative momentum spaces, as well as the linear entanglement entropy between them, converge to the same three-regime dynamics as in our model, apart from a transient stage whose duration depends on the contact interaction strength and the initial state. The investigation of the multi-rotor case is in progress.

\begin{acknowledgments}
We thank Zhixing Zou for helpful discussions and two anonymous referees for their valuable comments that helped improve the manuscript. This work is supported by the National Key R\&D Program of China (Grant No. 2023YFA1407100) and the Natural Science Foundation of China (Grant No. 12475038, No. 12247106, and No. 12247101).
\end{acknowledgments}


\begin{thebibliography}{43}%
\makeatletter
\providecommand \@ifxundefined [1]{%
 \@ifx{#1\undefined}
}%
\providecommand \@ifnum [1]{%
 \ifnum #1\expandafter \@firstoftwo
 \else \expandafter \@secondoftwo
 \fi
}%
\providecommand \@ifx [1]{%
 \ifx #1\expandafter \@firstoftwo
 \else \expandafter \@secondoftwo
 \fi
}%
\providecommand \natexlab [1]{#1}%
\providecommand \enquote  [1]{``#1''}%
\providecommand \bibnamefont  [1]{#1}%
\providecommand \bibfnamefont [1]{#1}%
\providecommand \citenamefont [1]{#1}%
\providecommand \href@noop [0]{\@secondoftwo}%
\providecommand \href [0]{\begingroup \@sanitize@url \@href}%
\providecommand \@href[1]{\@@startlink{#1}\@@href}%
\providecommand \@@href[1]{\endgroup#1\@@endlink}%
\providecommand \@sanitize@url [0]{\catcode `\\12\catcode `\$12\catcode
  `\&12\catcode `\#12\catcode `\^12\catcode `\_12\catcode `\%12\relax}%
\providecommand \@@startlink[1]{}%
\providecommand \@@endlink[0]{}%
\providecommand \url  [0]{\begingroup\@sanitize@url \@url }%
\providecommand \@url [1]{\endgroup\@href {#1}{\urlprefix }}%
\providecommand \urlprefix  [0]{URL }%
\providecommand \Eprint [0]{\href }%
\providecommand \doibase [0]{https://doi.org/}%
\providecommand \selectlanguage [0]{\@gobble}%
\providecommand \bibinfo  [0]{\@secondoftwo}%
\providecommand \bibfield  [0]{\@secondoftwo}%
\providecommand \translation [1]{[#1]}%
\providecommand \BibitemOpen [0]{}%
\providecommand \bibitemStop [0]{}%
\providecommand \bibitemNoStop [0]{.\EOS\space}%
\providecommand \EOS [0]{\spacefactor3000\relax}%
\providecommand \BibitemShut  [1]{\csname bibitem#1\endcsname}%
\let\auto@bib@innerbib\@empty
\bibitem [{\citenamefont {Abanin}\ \emph {et~al.}(2019)\citenamefont {Abanin},
  \citenamefont {Altman}, \citenamefont {Bloch},\ and\ \citenamefont
  {Serbyn}}]{Abanin2019}%
  \BibitemOpen
  \bibfield  {author} {\bibinfo {author} {\bibfnamefont {D.~A.}\ \bibnamefont
  {Abanin}}, \bibinfo {author} {\bibfnamefont {E.}~\bibnamefont {Altman}},
  \bibinfo {author} {\bibfnamefont {I.}~\bibnamefont {Bloch}},\ and\ \bibinfo
  {author} {\bibfnamefont {M.}~\bibnamefont {Serbyn}},\ }\bibfield  {title}
  {\bibinfo {title} {Colloquium: {{Many-body}} localization, thermalization,
  and entanglement},\ }\href {https://doi.org/10.1103/RevModPhys.91.021001}
  {\bibfield  {journal} {\bibinfo  {journal} {Rev. Mod. Phys.}\ }\textbf
  {\bibinfo {volume} {91}},\ \bibinfo {pages} {021001} (\bibinfo {year}
  {2019})}\BibitemShut {NoStop}%
\bibitem [{\citenamefont {Kaufman}\ \emph {et~al.}(2016)\citenamefont
  {Kaufman}, \citenamefont {Tai}, \citenamefont {Lukin}, \citenamefont
  {Rispoli}, \citenamefont {Schittko}, \citenamefont {Preiss},\ and\
  \citenamefont {Greiner}}]{Kaufman2016}%
  \BibitemOpen
  \bibfield  {author} {\bibinfo {author} {\bibfnamefont {A.~M.}\ \bibnamefont
  {Kaufman}}, \bibinfo {author} {\bibfnamefont {M.~E.}\ \bibnamefont {Tai}},
  \bibinfo {author} {\bibfnamefont {A.}~\bibnamefont {Lukin}}, \bibinfo
  {author} {\bibfnamefont {M.}~\bibnamefont {Rispoli}}, \bibinfo {author}
  {\bibfnamefont {R.}~\bibnamefont {Schittko}}, \bibinfo {author}
  {\bibfnamefont {P.~M.}\ \bibnamefont {Preiss}},\ and\ \bibinfo {author}
  {\bibfnamefont {M.}~\bibnamefont {Greiner}},\ }\bibfield  {title} {\bibinfo
  {title} {Quantum thermalization through entanglement in an isolated many-body
  system},\ }\href {https://doi.org/10.1126/science.aaf6725} {\bibfield
  {journal} {\bibinfo  {journal} {Science}\ }\textbf {\bibinfo {volume}
  {353}},\ \bibinfo {pages} {794} (\bibinfo {year} {2016})}\BibitemShut
  {NoStop}%
\bibitem [{\citenamefont {Jacquod}\ and\ \citenamefont
  {Petitjean}(2009)}]{Jacquod2009}%
  \BibitemOpen
  \bibfield  {author} {\bibinfo {author} {\bibfnamefont {P.}~\bibnamefont
  {Jacquod}}\ and\ \bibinfo {author} {\bibfnamefont {C.}~\bibnamefont
  {Petitjean}},\ }\bibfield  {title} {\bibinfo {title} {Decoherence,
  entanglement and irreversibility in quantum dynamical systems with few
  degrees of freedom},\ }\href {https://doi.org/10.1080/00018730902831009}
  {\bibfield  {journal} {\bibinfo  {journal} {Advances in Physics}\ }\textbf
  {\bibinfo {volume} {58}},\ \bibinfo {pages} {67} (\bibinfo {year}
  {2009})}\BibitemShut {NoStop}%
\bibitem [{\citenamefont {Calabrese}\ and\ \citenamefont
  {Cardy}(2005)}]{Calabrese2005}%
  \BibitemOpen
  \bibfield  {author} {\bibinfo {author} {\bibfnamefont {P.}~\bibnamefont
  {Calabrese}}\ and\ \bibinfo {author} {\bibfnamefont {J.}~\bibnamefont
  {Cardy}},\ }\bibfield  {title} {\bibinfo {title} {Evolution of entanglement
  entropy in one-dimensional systems},\ }\href
  {https://doi.org/10.1088/1742-5468/2005/04/P04010} {\bibfield  {journal}
  {\bibinfo  {journal} {J. Stat. Mech: Theory Exp.}\ }\textbf {\bibinfo
  {volume} {2005}},\ \bibinfo {pages} {P04010} (\bibinfo {year}
  {2005})}\BibitemShut {NoStop}%
\bibitem [{\citenamefont {Lerose}\ and\ \citenamefont
  {Pappalardi}(2020)}]{Lerose2020}%
  \BibitemOpen
  \bibfield  {author} {\bibinfo {author} {\bibfnamefont {A.}~\bibnamefont
  {Lerose}}\ and\ \bibinfo {author} {\bibfnamefont {S.}~\bibnamefont
  {Pappalardi}},\ }\bibfield  {title} {\bibinfo {title} {Bridging entanglement
  dynamics and chaos in semiclassical systems},\ }\href
  {https://doi.org/10.1103/PhysRevA.102.032404} {\bibfield  {journal} {\bibinfo
   {journal} {Phys. Rev. A}\ }\textbf {\bibinfo {volume} {102}},\ \bibinfo
  {pages} {032404} (\bibinfo {year} {2020})}\BibitemShut {NoStop}%
\bibitem [{\citenamefont {Lakshminarayan}\ \emph {et~al.}(2016)\citenamefont
  {Lakshminarayan}, \citenamefont {Srivastava}, \citenamefont {Ketzmerick},
  \citenamefont {B\"acker},\ and\ \citenamefont {Tomsovic}}]{Laksh2016}%
  \BibitemOpen
  \bibfield  {author} {\bibinfo {author} {\bibfnamefont {A.}~\bibnamefont
  {Lakshminarayan}}, \bibinfo {author} {\bibfnamefont {S.~C.~L.}\ \bibnamefont
  {Srivastava}}, \bibinfo {author} {\bibfnamefont {R.}~\bibnamefont
  {Ketzmerick}}, \bibinfo {author} {\bibfnamefont {A.}~\bibnamefont
  {B\"acker}},\ and\ \bibinfo {author} {\bibfnamefont {S.}~\bibnamefont
  {Tomsovic}},\ }\bibfield  {title} {\bibinfo {title} {Entanglement and
  localization transitions in eigenstates of interacting chaotic systems},\
  }\href {https://doi.org/10.1103/PhysRevE.94.010205} {\bibfield  {journal}
  {\bibinfo  {journal} {Phys. Rev. E}\ }\textbf {\bibinfo {volume} {94}},\
  \bibinfo {pages} {010205} (\bibinfo {year} {2016})}\BibitemShut {NoStop}%
\bibitem [{\citenamefont {Adachi}\ \emph {et~al.}(1988)\citenamefont {Adachi},
  \citenamefont {Toda},\ and\ \citenamefont {Ikeda}}]{Adachi1988}%
  \BibitemOpen
  \bibfield  {author} {\bibinfo {author} {\bibfnamefont {S.}~\bibnamefont
  {Adachi}}, \bibinfo {author} {\bibfnamefont {M.}~\bibnamefont {Toda}},\ and\
  \bibinfo {author} {\bibfnamefont {K.}~\bibnamefont {Ikeda}},\ }\bibfield
  {title} {\bibinfo {title} {Quantum-classical correspondence in
  many-dimensional quantum chaos},\ }\href
  {https://doi.org/10.1103/PhysRevLett.61.659} {\bibfield  {journal} {\bibinfo
  {journal} {Phys. Rev. Lett.}\ }\textbf {\bibinfo {volume} {61}},\ \bibinfo
  {pages} {659} (\bibinfo {year} {1988})}\BibitemShut {NoStop}%
\bibitem [{\citenamefont {Doron}\ and\ \citenamefont
  {Fishman}(1988)}]{Doron1988}%
  \BibitemOpen
  \bibfield  {author} {\bibinfo {author} {\bibfnamefont {E.}~\bibnamefont
  {Doron}}\ and\ \bibinfo {author} {\bibfnamefont {S.}~\bibnamefont
  {Fishman}},\ }\bibfield  {title} {\bibinfo {title} {Anderson localization for
  a two-dimensional rotor},\ }\href
  {https://doi.org/10.1103/PhysRevLett.60.867} {\bibfield  {journal} {\bibinfo
  {journal} {Phys. Rev. Lett.}\ }\textbf {\bibinfo {volume} {60}},\ \bibinfo
  {pages} {867} (\bibinfo {year} {1988})}\BibitemShut {NoStop}%
\bibitem [{\citenamefont {Gadway}\ \emph {et~al.}(2013)\citenamefont {Gadway},
  \citenamefont {Reeves}, \citenamefont {Krinner},\ and\ \citenamefont
  {Schneble}}]{Gadway2013}%
  \BibitemOpen
  \bibfield  {author} {\bibinfo {author} {\bibfnamefont {B.}~\bibnamefont
  {Gadway}}, \bibinfo {author} {\bibfnamefont {J.}~\bibnamefont {Reeves}},
  \bibinfo {author} {\bibfnamefont {L.}~\bibnamefont {Krinner}},\ and\ \bibinfo
  {author} {\bibfnamefont {D.}~\bibnamefont {Schneble}},\ }\bibfield  {title}
  {\bibinfo {title} {Evidence for a quantum-to-classical transition in a pair
  of coupled quantum rotors},\ }\href
  {https://doi.org/10.1103/PhysRevLett.110.190401} {\bibfield  {journal}
  {\bibinfo  {journal} {Phys. Rev. Lett.}\ }\textbf {\bibinfo {volume} {110}},\
  \bibinfo {pages} {190401} (\bibinfo {year} {2013})}\BibitemShut {NoStop}%
\bibitem [{\citenamefont {Matsui}\ \emph {et~al.}(2016)\citenamefont {Matsui},
  \citenamefont {Yamada},\ and\ \citenamefont {Ikeda}}]{Matsui2016}%
  \BibitemOpen
  \bibfield  {author} {\bibinfo {author} {\bibfnamefont {F.}~\bibnamefont
  {Matsui}}, \bibinfo {author} {\bibfnamefont {H.~S.}\ \bibnamefont {Yamada}},\
  and\ \bibinfo {author} {\bibfnamefont {K.~S.}\ \bibnamefont {Ikeda}},\
  }\bibfield  {title} {\bibinfo {title} {Relation between irreversibility and
  entanglement in classically chaotic quantum kicked rotors},\ }\href
  {https://doi.org/10.1209/0295-5075/114/60010} {\bibfield  {journal} {\bibinfo
   {journal} {Europhysics Letters}\ }\textbf {\bibinfo {volume} {114}},\
  \bibinfo {pages} {60010} (\bibinfo {year} {2016})}\BibitemShut {NoStop}%
\bibitem [{\citenamefont {Notarnicola}\ \emph {et~al.}(2018)\citenamefont
  {Notarnicola}, \citenamefont {Iemini}, \citenamefont {Rossini}, \citenamefont
  {Fazio}, \citenamefont {Silva},\ and\ \citenamefont
  {Russomanno}}]{Notar2018}%
  \BibitemOpen
  \bibfield  {author} {\bibinfo {author} {\bibfnamefont {S.}~\bibnamefont
  {Notarnicola}}, \bibinfo {author} {\bibfnamefont {F.}~\bibnamefont {Iemini}},
  \bibinfo {author} {\bibfnamefont {D.}~\bibnamefont {Rossini}}, \bibinfo
  {author} {\bibfnamefont {R.}~\bibnamefont {Fazio}}, \bibinfo {author}
  {\bibfnamefont {A.}~\bibnamefont {Silva}},\ and\ \bibinfo {author}
  {\bibfnamefont {A.}~\bibnamefont {Russomanno}},\ }\bibfield  {title}
  {\bibinfo {title} {From localization to anomalous diffusion in the dynamics
  of coupled kicked rotors},\ }\href
  {https://doi.org/10.1103/PhysRevE.97.022202} {\bibfield  {journal} {\bibinfo
  {journal} {Phys. Rev. E}\ }\textbf {\bibinfo {volume} {97}},\ \bibinfo
  {pages} {022202} (\bibinfo {year} {2018})}\BibitemShut {NoStop}%
\bibitem [{\citenamefont {Rozenbaum}\ and\ \citenamefont
  {Galitski}(2017)}]{Rozenbaum2017}%
  \BibitemOpen
  \bibfield  {author} {\bibinfo {author} {\bibfnamefont {E.~B.}\ \bibnamefont
  {Rozenbaum}}\ and\ \bibinfo {author} {\bibfnamefont {V.}~\bibnamefont
  {Galitski}},\ }\bibfield  {title} {\bibinfo {title} {Dynamical localization
  of coupled relativistic kicked rotors},\ }\href
  {https://doi.org/10.1103/PhysRevB.95.064303} {\bibfield  {journal} {\bibinfo
  {journal} {Phys. Rev. B}\ }\textbf {\bibinfo {volume} {95}},\ \bibinfo
  {pages} {064303} (\bibinfo {year} {2017})}\BibitemShut {NoStop}%
\bibitem [{\citenamefont {Paul}\ and\ \citenamefont {B\"acker}(2020)}]{SP2020}%
  \BibitemOpen
  \bibfield  {author} {\bibinfo {author} {\bibfnamefont {S.}~\bibnamefont
  {Paul}}\ and\ \bibinfo {author} {\bibfnamefont {A.}~\bibnamefont
  {B\"acker}},\ }\bibfield  {title} {\bibinfo {title} {Linear and logarithmic
  entanglement production in an interacting chaotic system},\ }\href
  {https://doi.org/10.1103/PhysRevE.102.050102} {\bibfield  {journal} {\bibinfo
   {journal} {Phys. Rev. E}\ }\textbf {\bibinfo {volume} {102}},\ \bibinfo
  {pages} {050102} (\bibinfo {year} {2020})}\BibitemShut {NoStop}%
\bibitem [{\citenamefont {Pulikkottil}\ \emph {et~al.}(2020)\citenamefont
  {Pulikkottil}, \citenamefont {Lakshminarayan}, \citenamefont {Srivastava},
  \citenamefont {B\"acker},\ and\ \citenamefont {Tomsovic}}]{JJP2020}%
  \BibitemOpen
  \bibfield  {author} {\bibinfo {author} {\bibfnamefont {J.~J.}\ \bibnamefont
  {Pulikkottil}}, \bibinfo {author} {\bibfnamefont {A.}~\bibnamefont
  {Lakshminarayan}}, \bibinfo {author} {\bibfnamefont {S.~C.~L.}\ \bibnamefont
  {Srivastava}}, \bibinfo {author} {\bibfnamefont {A.}~\bibnamefont
  {B\"acker}},\ and\ \bibinfo {author} {\bibfnamefont {S.}~\bibnamefont
  {Tomsovic}},\ }\bibfield  {title} {\bibinfo {title} {Entanglement production
  by interaction quenches of quantum chaotic subsystems},\ }\href
  {https://doi.org/10.1103/PhysRevE.101.032212} {\bibfield  {journal} {\bibinfo
   {journal} {Phys. Rev. E}\ }\textbf {\bibinfo {volume} {101}},\ \bibinfo
  {pages} {032212} (\bibinfo {year} {2020})}\BibitemShut {NoStop}%
\bibitem [{\citenamefont {Nambudiripad}\ \emph {et~al.}(2024)\citenamefont
  {Nambudiripad}, \citenamefont {Kannan},\ and\ \citenamefont
  {Santhanam}}]{AN2024}%
  \BibitemOpen
  \bibfield  {author} {\bibinfo {author} {\bibfnamefont {A.}~\bibnamefont
  {Nambudiripad}}, \bibinfo {author} {\bibfnamefont {J.~B.}\ \bibnamefont
  {Kannan}},\ and\ \bibinfo {author} {\bibfnamefont {M.~S.}\ \bibnamefont
  {Santhanam}},\ }\bibfield  {title} {\bibinfo {title} {Chaos and localized
  phases in a two-body linear kicked rotor system},\ }\href
  {https://doi.org/10.1103/PhysRevE.109.034206} {\bibfield  {journal} {\bibinfo
   {journal} {Phys. Rev. E}\ }\textbf {\bibinfo {volume} {109}},\ \bibinfo
  {pages} {034206} (\bibinfo {year} {2024})}\BibitemShut {NoStop}%
\bibitem [{\citenamefont {Fujisaki}\ \emph {et~al.}(2003)\citenamefont
  {Fujisaki}, \citenamefont {Tanaka},\ and\ \citenamefont
  {Miyadera}}]{Fujisaki2003}%
  \BibitemOpen
  \bibfield  {author} {\bibinfo {author} {\bibfnamefont {H.}~\bibnamefont
  {Fujisaki}}, \bibinfo {author} {\bibfnamefont {A.}~\bibnamefont {Tanaka}},\
  and\ \bibinfo {author} {\bibfnamefont {T.}~\bibnamefont {Miyadera}},\
  }\bibfield  {title} {\bibinfo {title} {Dynamical aspects of quantum
  entanglement for coupled mapping systems},\ }\href
  {https://doi.org/10.1143/JPSJS.72SC.111} {\bibfield  {journal} {\bibinfo
  {journal} {J. Phys. Soc. Jpn.}\ }\textbf {\bibinfo {volume} {72}},\ \bibinfo
  {pages} {111} (\bibinfo {year} {2003})}\BibitemShut {NoStop}%
\bibitem [{\citenamefont {Paul}\ \emph {et~al.}(2024)\citenamefont {Paul},
  \citenamefont {Kannan},\ and\ \citenamefont {Santhanam}}]{SP2024}%
  \BibitemOpen
  \bibfield  {author} {\bibinfo {author} {\bibfnamefont {S.}~\bibnamefont
  {Paul}}, \bibinfo {author} {\bibfnamefont {J.~B.}\ \bibnamefont {Kannan}},\
  and\ \bibinfo {author} {\bibfnamefont {M.~S.}\ \bibnamefont {Santhanam}},\
  }\bibfield  {title} {\bibinfo {title} {Faster entanglement driven by quantum
  resonance in many-body kicked rotors},\ }\href
  {https://doi.org/10.1103/PhysRevB.110.144301} {\bibfield  {journal} {\bibinfo
   {journal} {Phys. Rev. B}\ }\textbf {\bibinfo {volume} {110}},\ \bibinfo
  {pages} {144301} (\bibinfo {year} {2024})}\BibitemShut {NoStop}%
\bibitem [{\citenamefont {Casati}\ \emph {et~al.}(1979)\citenamefont {Casati},
  \citenamefont {Chirikov}, \citenamefont {Izraelev},\ and\ \citenamefont
  {Ford}}]{Casati1979}%
  \BibitemOpen
  \bibfield  {author} {\bibinfo {author} {\bibfnamefont {G.}~\bibnamefont
  {Casati}}, \bibinfo {author} {\bibfnamefont {B.~V.}\ \bibnamefont
  {Chirikov}}, \bibinfo {author} {\bibfnamefont {F.~M.}\ \bibnamefont
  {Izraelev}},\ and\ \bibinfo {author} {\bibfnamefont {J.}~\bibnamefont
  {Ford}},\ }\bibfield  {title} {\bibinfo {title} {Stochastic behavior of a
  quantum pendulum under a periodic perturbation},\ }in\ \href
  {https://doi.org/10.1007/BFb0021757} {\emph {\bibinfo {booktitle} {Stochastic
  Behavior in Classical and Quantum Hamiltonian Systems}}},\ \bibinfo {editor}
  {edited by\ \bibinfo {editor} {\bibfnamefont {G.}~\bibnamefont {Casati}}\
  and\ \bibinfo {editor} {\bibfnamefont {J.}~\bibnamefont {Ford}}}\ (\bibinfo
  {publisher} {Springer,  Berlin}, \ \bibinfo {year} {1979})\ pp.\ \bibinfo {pages}
  {334--352}\BibitemShut {NoStop}%
\bibitem [{\citenamefont {Fishman}\ \emph {et~al.}(1982)\citenamefont
  {Fishman}, \citenamefont {Grempel},\ and\ \citenamefont
  {Prange}}]{Fishman1982}%
  \BibitemOpen
  \bibfield  {author} {\bibinfo {author} {\bibfnamefont {S.}~\bibnamefont
  {Fishman}}, \bibinfo {author} {\bibfnamefont {D.~R.}\ \bibnamefont
  {Grempel}},\ and\ \bibinfo {author} {\bibfnamefont {R.~E.}\ \bibnamefont
  {Prange}},\ }\bibfield  {title} {\bibinfo {title} {Chaos, quantum
  recurrences, and anderson localization},\ }\href
  {https://doi.org/10.1103/PhysRevLett.49.509} {\bibfield  {journal} {\bibinfo
  {journal} {Phys. Rev. Lett.}\ }\textbf {\bibinfo {volume} {49}},\ \bibinfo
  {pages} {509} (\bibinfo {year} {1982})}\BibitemShut {NoStop}%
\bibitem [{\citenamefont {Izrailev}(1990)}]{Izrailev1990}%
  \BibitemOpen
  \bibfield  {author} {\bibinfo {author} {\bibfnamefont {F.~M.}\ \bibnamefont
  {Izrailev}},\ }\bibfield  {title} {\bibinfo {title} {Simple models of quantum
  chaos: Spectrum and eigenfunctions},\ }\href
  {https://doi.org/10.1016/0370-1573(90)90067-C} {\bibfield  {journal}
  {\bibinfo  {journal} {Phys. Rep.}\ }\textbf {\bibinfo {volume} {196}},\
  \bibinfo {pages} {299} (\bibinfo {year} {1990})}\BibitemShut {NoStop}%
\bibitem [{\citenamefont {Casati}\ and\ \citenamefont
  {Chirikov}(1995)}]{Casati1995}%
  \BibitemOpen
  \bibfield  {author} {\bibinfo {author} {\bibfnamefont {G.}~\bibnamefont
  {Casati}}\ and\ \bibinfo {author} {\bibfnamefont {B.}~\bibnamefont
  {Chirikov}},\ }\href@noop {} {\emph {\bibinfo {title} {Quantum Chaos:
  {{Between}} Order and Disorder}}}\ (\bibinfo  {publisher} {Cambridge
  University Press},\ \bibinfo {year} {1995})\BibitemShut {NoStop}%
\bibitem [{\citenamefont {Garreau}(2017)}]{Garreau2017}%
  \BibitemOpen
  \bibfield  {author} {\bibinfo {author} {\bibfnamefont {J.-C.}\ \bibnamefont
  {Garreau}},\ }\bibfield  {title} {\bibinfo {title} {Quantum simulation of
  disordered systems with cold atoms},\ }\href
  {https://doi.org/https://doi.org/10.1016/j.crhy.2016.09.002} {\bibfield
  {journal} {\bibinfo  {journal} {C. R. Physique}\ }\textbf {\bibinfo
  {volume} {18}},\ \bibinfo {pages} {31} (\bibinfo {year} {2017})}\BibitemShut {NoStop}%
\bibitem [{\citenamefont {Izrailev}\ and\ \citenamefont
  {Shepelyanskii}(1980)}]{Izrailev1980}%
  \BibitemOpen
  \bibfield  {author} {\bibinfo {author} {\bibfnamefont {F.~M.}\ \bibnamefont
  {Izrailev}}\ and\ \bibinfo {author} {\bibfnamefont {D.~L.}\ \bibnamefont
  {Shepelyanskii}},\ }\bibfield  {title} {\bibinfo {title} {Quantum resonance
  for a rotator in a nonlinear periodic field},\ }\href
  {https://doi.org/10.1007/BF01029131} {\bibfield  {journal} {\bibinfo
  {journal} {Theor. Math. Phys.}\ }\textbf {\bibinfo {volume} {43}},\ \bibinfo
  {pages} {553} (\bibinfo {year} {1980})}\BibitemShut {NoStop}%
\bibitem [{Note1()}]{Note1}%
  \BibitemOpen
  \bibinfo {note} {Applying the two operators $\exp ({-i{\pi p^2}})$ and $\exp
  ({-i{\pi p}})$ to an eigenstate $|l\rangle $ of $p$ leads to the same result,
  i.e., $\exp (-il\pi )|l\rangle =(-1)^l |l\rangle $, due to the fact that an
  integer and its square share the same parity. It follows straightforwardly
  that the two operators are equivalent.}\BibitemShut {Stop}%
\bibitem [{\citenamefont {Dana}\ \emph {et~al.}(1996)\citenamefont {Dana},
  \citenamefont {Eisenberg},\ and\ \citenamefont {Shnerb}}]{Dana1996}%
  \BibitemOpen
  \bibfield  {author} {\bibinfo {author} {\bibfnamefont {I.}~\bibnamefont
  {Dana}}, \bibinfo {author} {\bibfnamefont {E.}~\bibnamefont {Eisenberg}},\
  and\ \bibinfo {author} {\bibfnamefont {N.}~\bibnamefont {Shnerb}},\
  }\bibfield  {title} {\bibinfo {title} {Antiresonance and localization in
  quantum dynamics},\ }\href {https://doi.org/10.1103/PhysRevE.54.5948}
  {\bibfield  {journal} {\bibinfo  {journal} {Phys. Rev. E}\ }\textbf {\bibinfo
  {volume} {54}},\ \bibinfo {pages} {5948} (\bibinfo {year}
  {1996})}\BibitemShut {NoStop}%
\bibitem [{\citenamefont {Islam}\ \emph {et~al.}(2015)\citenamefont {Islam},
  \citenamefont {Ma}, \citenamefont {Preiss}, \citenamefont {Eric~Tai},
  \citenamefont {Lukin}, \citenamefont {Rispoli},\ and\ \citenamefont
  {Greiner}}]{Islam2015}%
  \BibitemOpen
  \bibfield  {author} {\bibinfo {author} {\bibfnamefont {R.}~\bibnamefont
  {Islam}}, \bibinfo {author} {\bibfnamefont {R.}~\bibnamefont {Ma}}, \bibinfo
  {author} {\bibfnamefont {P.~M.}\ \bibnamefont {Preiss}}, \bibinfo {author}
  {\bibfnamefont {M.}~\bibnamefont {Eric~Tai}}, \bibinfo {author}
  {\bibfnamefont {A.}~\bibnamefont {Lukin}}, \bibinfo {author} {\bibfnamefont
  {M.}~\bibnamefont {Rispoli}},\ and\ \bibinfo {author} {\bibfnamefont
  {M.}~\bibnamefont {Greiner}},\ }\bibfield  {title} {\bibinfo {title}
  {Measuring entanglement entropy in a quantum many-body system},\ }\href
  {https://doi.org/10.1038/nature15750} {\bibfield  {journal} {\bibinfo
  {journal} {Nature}\ }\textbf {\bibinfo {volume} {528}},\ \bibinfo {pages}
  {77} (\bibinfo {year} {2015})}\BibitemShut {NoStop}%
\bibitem [{\citenamefont {Ho}\ and\ \citenamefont {Abanin}(2017)}]{Ho2017}%
  \BibitemOpen
  \bibfield  {author} {\bibinfo {author} {\bibfnamefont {W.~W.}\ \bibnamefont
  {Ho}}\ and\ \bibinfo {author} {\bibfnamefont {D.~A.}\ \bibnamefont
  {Abanin}},\ }\bibfield  {title} {\bibinfo {title} {Entanglement dynamics in
  quantum many-body systems},\ }\href
  {https://doi.org/10.1103/PhysRevB.95.094302} {\bibfield  {journal} {\bibinfo
  {journal} {Phys. Rev. B}\ }\textbf {\bibinfo {volume} {95}},\ \bibinfo
  {pages} {094302} (\bibinfo {year} {2017})}\BibitemShut {NoStop}%
\bibitem [{\citenamefont {Wang}\ and\ \citenamefont
  {Zanardi}(2002)}]{Wang2002}%
  \BibitemOpen
  \bibfield  {author} {\bibinfo {author} {\bibfnamefont {X.}~\bibnamefont
  {Wang}}\ and\ \bibinfo {author} {\bibfnamefont {P.}~\bibnamefont {Zanardi}},\
  }\bibfield  {title} {\bibinfo {title} {Quantum entanglement of unitary
  operators on bipartite systems},\ }\href
  {https://doi.org/10.1103/PhysRevA.66.044303} {\bibfield  {journal} {\bibinfo
  {journal} {Phys. Rev. A}\ }\textbf {\bibinfo {volume} {66}},\ \bibinfo
  {pages} {044303} (\bibinfo {year} {2002})}\BibitemShut {NoStop}%
\bibitem [{\citenamefont {Sakurai}\ and\ \citenamefont
  {Napolitano}(2020)}]{Sakurai2020}%
  \BibitemOpen
  \bibfield  {author} {\bibinfo {author} {\bibfnamefont {J.~J.}\ \bibnamefont
  {Sakurai}}\ and\ \bibinfo {author} {\bibfnamefont {J.}~\bibnamefont
  {Napolitano}},\ }\href@noop {} {\emph {\bibinfo {title} {Modern Quantum
  Mechanics}}}\ (\bibinfo  {publisher} {Cambridge University Press, Cambridge},\ \bibinfo
  {year} {2020})\BibitemShut {NoStop}%
\bibitem [{\citenamefont {Russomanno}(2023)}]{Russomanno2023}%
  \BibitemOpen
  \bibfield  {author} {\bibinfo {author} {\bibfnamefont {A.}~\bibnamefont
  {Russomanno}},\ }\bibfield  {title} {\bibinfo {title} {Spatiotemporally
  ordered patterns in a chain of coupled dissipative kicked rotors},\ }\href
  {https://doi.org/10.1103/PhysRevB.108.094305} {\bibfield  {journal} {\bibinfo
   {journal} {Phys. Rev. B}\ }\textbf {\bibinfo {volume} {108}},\ \bibinfo
  {pages} {094305} (\bibinfo {year} {2023})}\BibitemShut {NoStop}%
\bibitem [{\citenamefont {Zou}\ and\ \citenamefont {Wang}(2024)}]{Zou2024}%
  \BibitemOpen
  \bibfield  {author} {\bibinfo {author} {\bibfnamefont {Z.}~\bibnamefont
  {Zou}}\ and\ \bibinfo {author} {\bibfnamefont {J.}~\bibnamefont {Wang}},\
  }\bibfield  {title} {\bibinfo {title} {A pseudoclassical theory for the
  wavepacket dynamics of the kicked rotor model},\ }\href
  {https://doi.org/10.1007/s11433-023-2279-6} {\bibfield  {journal} {\bibinfo
  {journal} {Sci. China: Phys. Mech. Astron.}\ }\textbf {\bibinfo {volume}
  {67}},\ \bibinfo {pages} {230511} (\bibinfo {year} {2024})}\BibitemShut
  {NoStop}%
\bibitem [{\citenamefont {Haake}\ \emph {et~al.}(1987)\citenamefont {Haake},
  \citenamefont {Ku{\'s}},\ and\ \citenamefont {Scharf}}]{Haake1987}%
  \BibitemOpen
  \bibfield  {author} {\bibinfo {author} {\bibfnamefont {F.}~\bibnamefont
  {Haake}}, \bibinfo {author} {\bibfnamefont {M.}~\bibnamefont {Ku{\'s}}},\
  and\ \bibinfo {author} {\bibfnamefont {R.}~\bibnamefont {Scharf}},\
  }\bibfield  {title} {\bibinfo {title} {Classical and quantum chaos for a
  kicked top},\ }\href@noop {} {\bibfield  {journal} {\bibinfo  {journal}
  {Zeitschrift f{\"u}r Physik B Condensed Matter}\ }\textbf {\bibinfo {volume}
  {65}},\ \bibinfo {pages} {381} (\bibinfo {year} {1987})}\BibitemShut
  {NoStop}%
\bibitem [{\citenamefont {Haake}\ and\ \citenamefont
  {Shepelyansky}(1988)}]{Haake1988}%
  \BibitemOpen
  \bibfield  {author} {\bibinfo {author} {\bibfnamefont {F.}~\bibnamefont
  {Haake}}\ and\ \bibinfo {author} {\bibfnamefont {D.~L.}\ \bibnamefont
  {Shepelyansky}},\ }\bibfield  {title} {\bibinfo {title} {The kicked rotator
  as a limit of the kicked top},\ }\href
  {https://doi.org/10.1209/0295-5075/5/8/001} {\bibfield  {journal} {\bibinfo
  {journal} {EPL}\ }\textbf {\bibinfo {volume} {5}},\ \bibinfo {pages} {671}
  (\bibinfo {year} {1988})}\BibitemShut {NoStop}%
\bibitem [{\citenamefont {Zou}\ and\ \citenamefont {Wang}(2022)}]{Zou2022}%
  \BibitemOpen
  \bibfield  {author} {\bibinfo {author} {\bibfnamefont {Z.}~\bibnamefont
  {Zou}}\ and\ \bibinfo {author} {\bibfnamefont {J.}~\bibnamefont {Wang}},\
  }\bibfield  {title} {\bibinfo {title} {Pseudoclassical dynamics of the kicked
  top},\ }\href {https://www.mdpi.com/1099-4300/24/8/1092} {\bibfield
  {journal} {\bibinfo  {journal} {Entropy}\ }\textbf {\bibinfo {volume} {24}}
  (\bibinfo {year} {2022})}\BibitemShut {NoStop}%
\bibitem [{\citenamefont {Parker}\ and\ \citenamefont
  {Stroud}(1986)}]{Parker1986}%
  \BibitemOpen
  \bibfield  {author} {\bibinfo {author} {\bibfnamefont {J.}~\bibnamefont
  {Parker}}\ and\ \bibinfo {author} {\bibfnamefont {C.~R.}\ \bibnamefont
  {Stroud}},\ }\bibfield  {title} {\bibinfo {title} {Coherence and decay of
  rydberg wave packets},\ }\href {https://doi.org/10.1103/PhysRevLett.56.716}
  {\bibfield  {journal} {\bibinfo  {journal} {Phys. Rev. Lett.}\ }\textbf
  {\bibinfo {volume} {56}},\ \bibinfo {pages} {716} (\bibinfo {year}
  {1986})}\BibitemShut {NoStop}%
\bibitem [{\citenamefont {Robinett}(2004)}]{Robinett2004}%
  \BibitemOpen
  \bibfield  {author} {\bibinfo {author} {\bibfnamefont {R.}~\bibnamefont
  {Robinett}},\ }\bibfield  {title} {\bibinfo {title} {Quantum wave packet
  revivals},\ }\href
  {https://doi.org/https://doi.org/10.1016/j.physrep.2003.11.002} {\bibfield
  {journal} {\bibinfo  {journal} {Physics Reports}\ }\textbf {\bibinfo {volume}
  {392}},\ \bibinfo {pages} {1} (\bibinfo {year} {2004})}\BibitemShut {NoStop}%
\bibitem [{\citenamefont {Rozenbaum}\ \emph {et~al.}(2017)\citenamefont
  {Rozenbaum}, \citenamefont {Ganeshan},\ and\ \citenamefont
  {Galitski}}]{Galitski2017}%
  \BibitemOpen
  \bibfield  {author} {\bibinfo {author} {\bibfnamefont {E.~B.}\ \bibnamefont
  {Rozenbaum}}, \bibinfo {author} {\bibfnamefont {S.}~\bibnamefont
  {Ganeshan}},\ and\ \bibinfo {author} {\bibfnamefont {V.}~\bibnamefont
  {Galitski}},\ }\bibfield  {title} {\bibinfo {title} {Lyapunov exponent and
  out-of-time-ordered correlator's growth rate in a chaotic system},\ }\href
  {https://doi.org/10.1103/PhysRevLett.118.086801} {\bibfield  {journal}
  {\bibinfo  {journal} {Phys. Rev. Lett.}\ }\textbf {\bibinfo {volume} {118}},\
  \bibinfo {pages} {086801} (\bibinfo {year} {2017})}\BibitemShut {NoStop}%
\bibitem [{\citenamefont {Haug}\ \emph {et~al.}(2005)\citenamefont {Haug},
  \citenamefont {Bienert}, \citenamefont {Schleich}, \citenamefont {Seligman},\
  and\ \citenamefont {Raizen}}]{Raizen2005}%
  \BibitemOpen
  \bibfield  {author} {\bibinfo {author} {\bibfnamefont {F.}~\bibnamefont
  {Haug}}, \bibinfo {author} {\bibfnamefont {M.}~\bibnamefont {Bienert}},
  \bibinfo {author} {\bibfnamefont {W.~P.}\ \bibnamefont {Schleich}}, \bibinfo
  {author} {\bibfnamefont {T.~H.}\ \bibnamefont {Seligman}},\ and\ \bibinfo
  {author} {\bibfnamefont {M.~G.}\ \bibnamefont {Raizen}},\ }\bibfield  {title}
  {\bibinfo {title} {Motional stability of the quantum kicked rotor: A fidelity
  approach},\ }\href {https://doi.org/10.1103/PhysRevA.71.043803} {\bibfield
  {journal} {\bibinfo  {journal} {Phys. Rev. A}\ }\textbf {\bibinfo {volume}
  {71}},\ \bibinfo {pages} {043803} (\bibinfo {year} {2005})}\BibitemShut
  {NoStop}%
\bibitem [{\citenamefont {Longhi}(2017)}]{Longhi2017}%
  \BibitemOpen
  \bibfield  {author} {\bibinfo {author} {\bibfnamefont {S.}~\bibnamefont
  {Longhi}},\ }\bibfield  {title} {\bibinfo {title} {Localization, quantum
  resonances, and ratchet acceleration in a periodically kicked
  $\mathcal{PT}$-symmetric quantum rotator},\ }\href
  {https://doi.org/10.1103/PhysRevA.95.012125} {\bibfield  {journal} {\bibinfo
  {journal} {Phys. Rev. A}\ }\textbf {\bibinfo {volume} {95}},\ \bibinfo
  {pages} {012125} (\bibinfo {year} {2017})}\BibitemShut {NoStop}%
\bibitem [{\citenamefont {Zhao}\ \emph {et~al.}(2019)\citenamefont {Zhao},
  \citenamefont {Wang}, \citenamefont {Wang},\ and\ \citenamefont
  {Tong}}]{Zhao2019}%
  \BibitemOpen
  \bibfield  {author} {\bibinfo {author} {\bibfnamefont {W.-L.}\ \bibnamefont
  {Zhao}}, \bibinfo {author} {\bibfnamefont {J.}~\bibnamefont {Wang}}, \bibinfo
  {author} {\bibfnamefont {X.}~\bibnamefont {Wang}},\ and\ \bibinfo {author}
  {\bibfnamefont {P.}~\bibnamefont {Tong}},\ }\bibfield  {title} {\bibinfo
  {title} {Directed momentum current induced by the $\mathcal{PT}$-symmetric
  driving},\ }\href {https://doi.org/10.1103/PhysRevE.99.042201} {\bibfield
  {journal} {\bibinfo  {journal} {Phys. Rev. E}\ }\textbf {\bibinfo {volume}
  {99}},\ \bibinfo {pages} {042201} (\bibinfo {year} {2019})}\BibitemShut
  {NoStop}%
\bibitem [{\citenamefont {Chen}\ \emph {et~al.}(2025)\citenamefont {Chen},
  \citenamefont {Ni}, \citenamefont {Song}, \citenamefont {Huang},
  \citenamefont {Wang},\ and\ \citenamefont {Casati}}]{Huang2025}%
  \BibitemOpen
  \bibfield  {author} {\bibinfo {author} {\bibfnamefont {Z.-Q.}\ \bibnamefont
  {Chen}}, \bibinfo {author} {\bibfnamefont {R.-H.}\ \bibnamefont {Ni}},
  \bibinfo {author} {\bibfnamefont {Y.}~\bibnamefont {Song}}, \bibinfo {author}
  {\bibfnamefont {L.}~\bibnamefont {Huang}}, \bibinfo {author} {\bibfnamefont
  {J.}~\bibnamefont {Wang}},\ and\ \bibinfo {author} {\bibfnamefont
  {G.}~\bibnamefont {Casati}},\ }\bibfield  {title} {\bibinfo {title}
  {Correspondence principle, ergodicity, and finite-time dynamics},\ }\href
  {https://doi.org/10.1103/PhysRevLett.134.130402} {\bibfield  {journal}
  {\bibinfo  {journal} {Phys. Rev. Lett.}\ }\textbf {\bibinfo {volume} {134}},\
  \bibinfo {pages} {130402} (\bibinfo {year} {2025})}\BibitemShut {NoStop}%
\bibitem [{\citenamefont {Guo}\ \emph {et~al.}(2025)\citenamefont {Guo},
  \citenamefont {Dhar}, \citenamefont {Yang}, \citenamefont {Chen},
  \citenamefont {Yao}, \citenamefont {Horvath}, \citenamefont {Ying},
  \citenamefont {Landini},\ and\ \citenamefont {Naegerl}}]{Ying2025}%
  \BibitemOpen
  \bibfield  {author} {\bibinfo {author} {\bibfnamefont {Y.}~\bibnamefont
  {Guo}}, \bibinfo {author} {\bibfnamefont {S.}~\bibnamefont {Dhar}}, \bibinfo
  {author} {\bibfnamefont {A.}~\bibnamefont {Yang}}, \bibinfo {author}
  {\bibfnamefont {Z.}~\bibnamefont {Chen}}, \bibinfo {author} {\bibfnamefont
  {H.}~\bibnamefont {Yao}}, \bibinfo {author} {\bibfnamefont {M.}~\bibnamefont
  {Horvath}}, \bibinfo {author} {\bibfnamefont {L.}~\bibnamefont {Ying}},
  \bibinfo {author} {\bibfnamefont {M.}~\bibnamefont {Landini}},\ and\ \bibinfo
  {author} {\bibfnamefont {H.-C.}\ \bibnamefont {Naegerl}},\ }\bibfield
  {title} {\bibinfo {title} {Observation of many-body dynamical localization},\
  }\href {https://doi.org/10.1126/science.adn8625} {\bibfield  {journal}
  {\bibinfo  {journal} {Science}\ }\textbf {\bibinfo {volume} {389}},\ \bibinfo
  {pages} {716} (\bibinfo {year} {2025})}\BibitemShut {NoStop}%
\bibitem [{\citenamefont {Qin}\ \emph {et~al.}(2017)\citenamefont {Qin},
  \citenamefont {Andreanov}, \citenamefont {Park},\ and\ \citenamefont
  {Flach}}]{Flach2017}%
  \BibitemOpen
  \bibfield  {author} {\bibinfo {author} {\bibfnamefont {P.}~\bibnamefont
  {Qin}}, \bibinfo {author} {\bibfnamefont {A.}~\bibnamefont {Andreanov}},
  \bibinfo {author} {\bibfnamefont {H.~C.}\ \bibnamefont {Park}},\ and\
  \bibinfo {author} {\bibfnamefont {S.}~\bibnamefont {Flach}},\ }\bibfield
  {title} {\bibinfo {title} {Interacting ultracold atomic kicked rotors: loss
  of dynamical localization},\ }\href {https://doi.org/10.1038/srep41139}
  {\bibfield  {journal} {\bibinfo  {journal} {Sci. Rep.}\ }\textbf {\bibinfo
  {volume} {7}},\ \bibinfo {pages} {41139} (\bibinfo {year}
  {2017})}\BibitemShut {NoStop}%
\end{thebibliography}

%

\widetext
\appendix

\section{Derivation of Eq.~(\ref{SlintDelta})}

The $t$-step evolution operator of our many-body kicked rotor model in the two lowest-order resonances is \(U^t=U^t_{\text{local}}U^t_I\), where \(U^t_I\) denotes the contribution from the interaction potential that couples the two subsystems, $A$ and $B$, the linear entanglement entropy between which is of interest here. As entanglement arises solely from the interaction, the linear entanglement entropy at time $t$ evaluated over the state  \(|\psi (t)\rangle=U^t|\psi(0)\rangle\) is identical to that over \(|\psi_I (t)\rangle=U_I^t|\psi(0)\rangle\). Hence we consider the latter for simplicity. Note that \(U^t_I=\exp(-i\mathcal{V}_{t,I})\), where the time-accumulated potential \(\mathcal{V}_{t,I}\) is defined in Eq.~\eqref{VST1}.

{\color{blue} For an arbitrary initial product state represented in the position basis as}
\begin{equation}
    |\psi(0)\rangle = \left (\int d\boldsymbol{\theta}_A \phi_A(\boldsymbol{\theta}_A)|\boldsymbol{\theta}_A\rangle\right )\otimes\left (  \int d\boldsymbol{\theta}_B\phi_B(\boldsymbol{\theta}_B)  |\boldsymbol{\theta}_B\rangle\right ),
\end{equation}
where \(|\boldsymbol{\theta}_A\rangle\) and \(|\boldsymbol{\theta}_B\rangle\) are the position eigenstates of subsystem  \(A\) and  \(B\), respectively, the evolved state at time $t$ in the interaction picture is
\begin{equation}
|\psi_I(t)\rangle = \int d\boldsymbol{\theta}_A d\boldsymbol{\theta}_B e^{-i\mathcal{V}_{t,I}(\boldsymbol{\theta}_A, \boldsymbol{\theta}_B)} \phi_A(\boldsymbol{\theta}_A)\phi_B(\boldsymbol{\theta}_B) |\boldsymbol{\theta}_A \boldsymbol{\theta}_B\rangle
\end{equation}
and the corresponding density operator ${\rho}_I (t)= |\psi_I(t)\rangle \langle \psi_I(t)|$ reads 
\begin{equation}
{\rho}_I (t) = \int d\boldsymbol{\theta}_A d\boldsymbol{\theta}_B d\boldsymbol{\theta}_{A'} d\boldsymbol{\theta}_{B'}~~e^{-i(\mathcal{V}_{t,I}(\boldsymbol{\theta}_A, \boldsymbol{\theta}_B)-\mathcal{V}_{t,I}(\boldsymbol{\theta}_{A'}, \boldsymbol{\theta}_{B'}))} \phi_A(\boldsymbol{\theta}_A)\phi_B(\boldsymbol{\theta}_B)\phi_A^*(\boldsymbol{\theta}_{A'})\phi_B^*(\boldsymbol{\theta}_{B'}) |\boldsymbol{\theta}_A\boldsymbol{\theta}_B\rangle \langle \boldsymbol{\theta}_{A'}\boldsymbol{\theta}_{B'}|.
\end{equation}
Tracing out subsystem $B$ yields the effective reduced density matrix for subsystem $A$:
\begin{equation}
\begin{aligned}
{\rho}_{I, A}(t) &= \mathrm{Tr}_B[{\rho}_I (t)]\\
&= \int d\boldsymbol{\theta}_A d\boldsymbol{\theta}_B d\boldsymbol{\theta}_{A'} ~~e^{-i(\mathcal{V}_{t,I}(\boldsymbol{\theta}_A,\boldsymbol{\theta}_B)-\mathcal{V}_{t,I}(\boldsymbol{\theta}_{A'},\boldsymbol{\theta}_B))} \phi_A(\boldsymbol{\theta}_A)\phi_A^*(\boldsymbol{\theta}_{A'}) \left|\phi_B(\boldsymbol{\theta}_B)\right|^2 |\boldsymbol{\theta}_A\rangle \langle \boldsymbol{\theta}_{A'}|.
\end{aligned}
\end{equation}
The purity can then be expressed as
\begin{equation}
\begin{aligned}
\mu_2(t) &= \mathrm{Tr}[{\rho}_{I,A}^2(t)] \\
&= \int d\boldsymbol{\theta}_A d\boldsymbol{\theta}_B d\boldsymbol{\theta}_{A'} d\boldsymbol{\theta}_{B'} \left|\phi_A(\boldsymbol{\theta}_A)\right|^2 \left|\phi_B(\boldsymbol{\theta}_B)\right|^2 \left|\phi_A(\boldsymbol{\theta}_{A'})\right|^2 \left|\phi_B(\boldsymbol{\theta}_{B'})\right|^2 e^{-i\Delta_t(\boldsymbol{\theta}_A ,\boldsymbol{\theta}_B,\boldsymbol{\theta}_{A'},\boldsymbol{\theta}_{B'})}\\
&=\left\langle \exp(-i\Delta_t)\right\rangle \\
&=\left\langle \cos(\Delta_t)\right\rangle,
\end{aligned}
\end{equation}
where $\langle \cdot \rangle$ represents the weighted average over the probability distribution of the initial state and \(\Delta_t(\boldsymbol{\theta}_A, \boldsymbol{\theta}_B, \boldsymbol{\theta}_{A'}, \boldsymbol{\theta}_{B'}) \equiv \mathcal{V}_{t,I}(\boldsymbol{\theta}_A, \boldsymbol{\theta}_B) + \mathcal{V}_{t,I}(\boldsymbol{\theta}_{A'}, \boldsymbol{\theta}_{B'})- \mathcal{V}_{t,I}(\boldsymbol{\theta}_{A'}, \boldsymbol{\theta}_B) - \mathcal{V}_{t,I}(\boldsymbol{\theta}_A, \boldsymbol{\theta}_{B'})\). The last equality comes from the fact that the purity must be real.  Specifically, we have
\begin{equation}
\begin{aligned}
\left\langle \cos(\Delta_t)\right\rangle =\int   d\boldsymbol{\theta}_A  d\boldsymbol{\theta}_B   d\boldsymbol{\theta}_{A'}  d\boldsymbol{\theta}_{B'}
\left|\phi_A(\boldsymbol{\theta}_A)\right|^2
\left |\phi_B(\boldsymbol{\theta}_B)\right|^2
\left|\phi_A(\boldsymbol{\theta}_{A'})\right|^2
\left |\phi_B(\boldsymbol{\theta}_{B'})\right|^2
\cos(\Delta_t(\boldsymbol{\theta}_A ,\boldsymbol{\theta}_B,\boldsymbol{\theta}_{A'},\boldsymbol{\theta}_{B'})).
\label{average}
\end{aligned}
\end{equation}
Note that  \(\left\langle \epsilon \right\rangle\) and  \(\left\langle \epsilon_\pm \right\rangle\) are defined in the same form and can be derived via Eq. \eqref{average} by simply replacing \(\cos(\Delta_t)\) with \(\epsilon\) and \(\epsilon_\pm\), respectively.

Finally, by definition, we get the linear entropy,
\begin{equation}
    \begin{aligned}
        S_{\text{lin}}(t) &= 1 - \mu_2(t) \\
        &= 1-\left\langle \cos(\Delta_t)\right\rangle,
        \label{slinapp}
    \end{aligned}
\end{equation}
i.e., Eq.~\eqref{SlintDelta} in the main text. 

\section{Entanglement generation for evolution operators with product eigenstates}
Here we show that the linear entropy generally exhibits an initial quadratic growth for any evolution operator $U$ that possesses a basis of tensor-product eigenstates. Suppose the eigenstates of $U$ are
\begin{equation}
	|ab\rangle = |a\rangle\otimes|b\rangle,
	\label{producteigen}
\end{equation}
where $\{|a\rangle, a = 1,2,...,d_{H_A}\}$ forms an orthonormal basis for subsystem \(A\) with Hilbert space dimension $d_{H_A}$ and $\{|b\rangle, b = 1,2,...,d_{H_B}\}$ for subsystem \(B\) of dimension $d_{H_B}$. The quasienergy $E_{ab}$ satisfies that
\begin{equation}
    U|ab\rangle = e^{-i E_{ab}}|ab\rangle.
\end{equation}
We consider a product initial state $|\psi(0)\rangle$ with zero entanglement entropy expanded as
\begin{equation}
\begin{aligned}
           |\psi(0)\rangle
   & = \Bigl(\sum_a \phi_a |a\rangle\Bigr)\otimes\Bigl(\sum_b \chi_b |b\rangle\Bigr)\\
    &\equiv \sum_{a,b}\varphi_{ab}\,|ab\rangle.
\end{aligned}
\end{equation}
After $t$ steps of evolution, the reduced density matrix for subsystem \(A\) is
\begin{equation}
    \begin{aligned}
        {\rho}_A(t) &= \mathrm{Tr}_B[{\rho}(t)] \\
        &= \sum_{ab,a'} \varphi_{ab}e^{-iE_{ab}t} |a\rangle \langle a'| \varphi_{a'b}^* e^{iE_{a'b}t},
    \end{aligned}
\end{equation}
leading to the purity
\begin{equation}
    \begin{aligned}
        \mu_2(t) &= \mathrm{Tr}[{\rho}_A^2(t)] \\
        &= \sum_{aba'b'} |\phi_a|^2|\chi_b|^2|\phi_{a'}|^2|\chi_{b'}|^2 e^{-i\epsilon_{aba'b'}t} \\
        &= \sum_{aba'b'} |\phi_a|^2|\chi_b|^2|\phi_{a'}|^2|\chi_{b'}|^2 \cos (\epsilon_{aba'b'}t),
    \end{aligned}
\end{equation}
where $\epsilon_{aba'b'}\equiv E_{ab}+E_{a'b'}-E_{a'b}-E_{ab'}$. The last equality is due to that $\mu_2(t)$ must be a real number. Setting
\begin{equation}
       \langle\epsilon^2\rangle \equiv \sum_{aba'b'} |\phi_a|^2|\chi_b|^2|\phi_{a'}|^2|\chi_{b'}|^2 \epsilon_{aba'b'}^2,
\end{equation}
a short-time expansion of \(\mu_2(t)\) in the regime \(t\ll t^*\sim 1/\sqrt{\langle\epsilon^2\rangle}\) gives
\begin{equation}
    \mu_2(t) \approx 1-\frac{\langle\epsilon^2\rangle}{2} t^2.
\end{equation}
With the definition of linear entropy $S_{\text{lin}}(t)=1 - \mu_2(t)$, we readily obtain that
\begin{equation}
    \begin{aligned}
        S_{\text{lin}}(t) \approx \frac{\langle\epsilon^2\rangle}{2} t^2.
        \label{eq:Slin_sinsq}
    \end{aligned}
\end{equation}

This shows that,  for any evolution operator with tensor-product eigenstates, the linear entanglement entropy exhibits universal quadratic initial growth from a generic separable initial state.

\end{document}